\documentclass[pra,twocolumn,english,aps,eqsecnum,showpacs,superscriptaddress]{revtex4-1}
\usepackage[T1]{fontenc}
\usepackage[latin9]{inputenc}
\usepackage{amsmath}
\usepackage{amssymb}
\usepackage{graphicx}

\makeatletter

\@ifundefined{textcolor}{}
{%
 \definecolor{BLACK}{gray}{0}
 \definecolor{WHITE}{gray}{1}
 \definecolor{RED}{rgb}{1,0,0}
 \definecolor{GREEN}{rgb}{0,1,0}
 \definecolor{BLUE}{rgb}{0,0,1}
 \definecolor{CYAN}{cmyk}{1,0,0,0}
 \definecolor{MAGENTA}{cmyk}{0,1,0,0}
 \definecolor{YELLOW}{cmyk}{0,0,1,0}
}

\usepackage{amsfonts}
\@ifundefined{definecolor}
 {\usepackage{color}}{}

\def\be{\begin{equation}}
\def\ee{\end{equation}}
\def\bea{\begin{eqnarray}}
\def\eea{\end{eqnarray}}
\def\bi{\begin{itemize}}
\def\ei{\end{itemize}}

\makeatother

\usepackage{babel}
\begin{document}

\title{One-dimensional frustrated plaquette compass model:\\
 Nematic phase and spontaneous multimerization}

\author{Wojciech Brzezicki}

\affiliation{Marian Smoluchowski Institute of Physics, Jagiellonian University,
prof. S. \L{}ojasiewicza 11, PL-30348 Krak\'ow, Poland }

\affiliation{CNR-SPIN, I-84084 Fisciano (SA), Italy, and \\
 Dipartimento di Fisica ``E. R. Caianiello\textquotedblright , Universit\'a
di Salerno, I-84084 Fisciano (SA), Italy}

\author{Andrzej M. Ole\'{s} }

\affiliation{Marian Smoluchowski Institute of Physics, Jagiellonian University,
prof. S. \L{}ojasiewicza 11, PL-30348 Krak\'ow, Poland }

\affiliation{Max-Planck-Institut f\"ur Festk\"orperforschung, Heisenbergstrasse 1,
D-70569 Stuttgart, Germany}

\date{16 October 2015}

\begin{abstract}
We introduce a one-dimensional (1D) pseudospin model on a ladder where
the Ising interactions along the legs and along the rungs alternate
between $X_{i}X_{i+1}$ and $Z_{i}Z_{i+1}$ for even/odd bond (rung).
We include also the next nearest neighbor Ising interactions on
plaquettes' diagonals that alternate in such a way that a model
where only leg interactions are switched on is equivalent to the one
when only the diagonal ones are present. Thus in the absence of rung
interactions the model can interpolate between two 1D compass models.
The model posses local symmetries which are the parities within each
$2\times 2$ cell (plaquette) of the ladder. We find that for
different values of the interaction it can realize ground states that
differ by the patterns formed by these local parities.
By exact diagonalization we derive detailed phase diagrams for small
systems of $L=4$, 6 and 8 plaquettes, and use next $L=12$ to identify
generic phases that appear in larger systems as well. Among them we
find a nematic phase with macroscopic degeneracy when the leg and
diagonal interactions are equal and the rung interactions are larger
than a critical value. By performing a perturbative expansion around
this phase we find indeed a very complex competition around the
nematic phase which has to do with releasing frustration in this
range of parameters. The nematic phase is similar to the one found in
the two-dimensional compass model. For particular parameters the
low-energy sector of the present plaquette model reduces to a 1D
compass model with spins $S=1$ which suggests that it realizes
peculiar crossovers within the class of compass models.
Finally, we show that the model can realize phases with broken
translation invariance which can be either dimerized, trimerized,
\textit{etcetera}, or completely disordered and highly entangled
in a~well identified window of the phase diagram.
\end{abstract}


\maketitle

\section{Introduction}
\label{sec:intro}

Entanglement in spin models is one of the central topics in modern
condensed matter theory \cite{Ami08}. It is frequently accompanied by
frustration of exchange interactions \cite{Nor09,Bal10}. However,
frustration alone is not sufficient to generate entanglement but in
many cases it triggers entangled excited or ground states. Some models
without entanglement are exactly solvable in two dimensions, as for
instance the fully frustrated two-dimensional (2D) Ising model of
Villain \cite{Vil77} with the reversed sign of exchange interaction
along every second column with respect to the unfrustrated square
lattice, and its generalization with periodically distributed
frustrated bonds along more distant columns \cite{Lon80}.

In contrast, the 2D compass model \cite{Nus15}, with competing $X_iX_j$
and $Z_iZ_j$ interactions between $S=1/2$ pseudospins ($X_i$ and $Z_i$
are Pauli matrices) along horizontal and vertical bonds in the square
lattice, is quantum and has intrinsic entanglement which can be reliably
treated only by advanced many-body methods \cite{Che07,Oru09}, including
quantum Monte Carlo \cite{Wen08}, multi-configurational entanglement
renormalization ansatz \cite{Cin10}, or tensor networks at finite
temperature \cite{Cza16}. The latter approach allowed to confirm that
long-range order develops in the 2D quantum compass model at finite
temperature, in analogy to the 2D classical Ising model, without or
with frustrated interactions \cite{Vil77,Lon80}. Moreover, the symmetry
properties of the 2D compass model are responsible for certain
relations between the correlations functions which may be viewed as
hidden order \cite{You10,Brz10}.

Recent interest in the compass models is motivated by several
developments:
(i) investigating quantum phase transitions \cite{Che07,Oru09,Cin10};
(ii) its relation to $p+ip$ superconductivity \cite{Nus05};
(iii) recently confirmed prospect of topological quantum computing
\cite{Kit03,Dou05,Dor05,Dus09} --- it could be realized in nanoscopic
systems where perturbing Heisenberg interactions do not destroy the
nematic order in the lowest energy excited states \cite{Tro10} while the
ground state remains ordered also at finite temperature \cite{Pla15};
(iv) order and excitations in case of compass interactions on
a frustrated checkerboard lattice \cite{Nas12};
(v) spin-orbital physics in transition metal oxides with active
orbital degrees of freedom
\cite{Kug82,Fei97,Hfm,Tokura,Kha01,Ver04,Ole05,Kha05,Ulr06,Hor08,
      Kru09,Karlo,Woh11,Ole12,Brz12,Cza15,Ave13,Brz15};
(vi) its realization in iridates with strong spin-orbit coupling
\cite{Jac09}, leading to the compass interactions on the triangular
lattice \cite{Jac15} or to the exactly solvable Kitaev model on the
honeycomb lattice \cite{Kit06,Bas07}, and
(vii)~its experimental realizations in optical lattices \cite{Mil07}.
In case of ferromagnetic (FM) spin-orbital systems entanglement is
absent \cite{Ole06} and one finds here quantum orbital models
\cite{vdB04,vdB99,Fei05,Ish05,Dag08,Ryn10,Wro10,Dag10,Tro13,Che13}.
The 2D compass model is their generic representation and its
interactions stand for directional orbital interactions between $e_g$
or $t_{2g}$ orbitals on the 2D square or on three-dimensional (3D)
cubic lattice. While both $e_g$ and $t_{2g}$
orbital models are distinct, none of them reduces to the compass
model \cite{Nus15,Wen11} which is more universal and stands for the
paradigm of directional interactions in the orbital physics
\cite{vdB99,vdB04,Fei05,Ish05,Dag08,Ryn10,Wro10,Dag10,Tro13,Che13}.

In spite of their conceptual simplicity, only very few quantum orbital
models are exactly solvable. Among them the Kitaev model on the honeycomb
lattice is the most prominent one as it describes a spin liquid with
only nearest neighbor (NN) spin correlations \cite{Kit06,Bas07}.
Recently considerable attention attracted also the one-dimensional
(1D) compass model which is exactly solvable by the mapping on the
transverse Ising model \cite{Brz07}. As the 2D compass model,
it includes only two spin components and the ground state is highly
degenerate. This triggers various phase transitions when the interactions
are tuned \cite{Sun09}. By investigating the block entanglement entropy
in the four ground-state phases it has been found that the changes
of entanglement signal the second-order rather than the first-order
transitions \cite{Eri09}. Further insights into the mechanism of
quantum phase transitions were obtained by matrix product state approach
\cite{Liu12,Wan15}. Furthermore, the exact solution was generalized
to the case with a finite transverse field \cite{Jaf11} which destabilizes
the orbital-liquid ground state with macroscopic degeneracy and rather
peculiar specific heat and polarization behavior of the 1D compass
model follows from highly frustrated interactions \cite{You14}.

Another exactly solvable case is the compass model on a ladder with
leg and rung interactions satisfying the same directional pattern
as in the 2D model \cite{Brz09}. In contrast, the 1D plaquette orbital
model with a topology of a ladder introduced recently \cite{Brz14}
is not exactly solvable but transforms to an effective 1D spin model
in a magnetic field, with spin dimers that replace plaquettes and
are coupled along the chain by three-spin interactions. This model
is motivated by the 2D plaquette orbital model \cite{Wen09,Bis10}
and has very interesting properties as the quantum effects are of
purely short-range nature which makes it possible to estimate the
ground state energy in the thermodynamic limit from the exact
diagonalization of finite clusters.

In the present paper we investigate a generalization of the 1D plaquette
compass model \cite{Brz14} with the diagonal pseudospin interactions
added within each plaquette. We present the phase diagrams obtained
by exact diagonalization for finite systems with periodic boundary
conditions (PBCs) which contain generic phases that are expected to
appear for any system size. This model interpolates between two 1D
compass models in absence of rung interactions. As we show below, the
nematic phase found in this frustrated plaquette orbital model is
similar to the one established for the 2D case \cite{Wen09} which
indicates that the present model realizes the paradigm of dimensional
crossover within the class of compass models.

The paper is organized as follows: in Sec. \ref{sec:hami} we introduce
the frustrated $Cx$-$Cz$ Hamiltonian and derive its block-diagonal form
making use of its local symmetries. Its symmetry line in the parameter
space is explored in Sec. \ref{sec:line}. Next we present a competition
between different classical states for a single plaquette in Sec.
\ref{sec:box} and show that this can be used as a guideline to
understand the complex quantum phase diagram for a generic case of $L$
interacting plaquettes, as shown for $L=4$ in Sec. \ref{sec:PD4} where
we also visualize the configurations found there. In Sec.
\ref{sec:aniso} we present the phase diagram for $L=4$ in the
anisotropic case when all the ZZ interaction are slightly weaker than
the XX ones. Phase diagrams for larger systems with $L>4$ are
investigated in Sec. \ref{sec:PhD2} --- in Sec. \ref{sec:PD6} we show
the detailed phase diagram for a ladder consisting of $L=$6 plaquettes
and its configurations, while Secs. \ref{sec:PD8} and \ref{sec:PD12}
concentrate on the evolution of phase diagram with increasing $L$ and
identification of generic phases using large ladders with $L=8$ and
$L=12$. Section \ref{sec:pert_nema} is devoted to phase competition in
the vicinity of the nematic phase and we study the behavior of the
energy levels when this phase is approached using perturbative
expansion up to fourth order. Finally, in Sec. \ref{sub:dim-dim} we
quantify the entanglement of the effective dimers in symmetry
subspaces for the systems of the sizes $L=4$ and $L=6$. In Sec.
\ref{sub:plaq-plaq} we do the same for plaquettes in initial physical
basis. The summary and conclusions are given in Sec. \ref{sec:summa}.
The paper is supplemented with one Appendix in which we show the
relation between the spin transformation that we use to
block-diagonalize the present model and the one which was used before
for a simpler $Cx$--$Cz$ model \cite{Brz14}.

\section{Hamilonian and its symmetries}
\label{sec:hami}

\begin{figure}[t!]
\includegraphics[width=1\columnwidth]{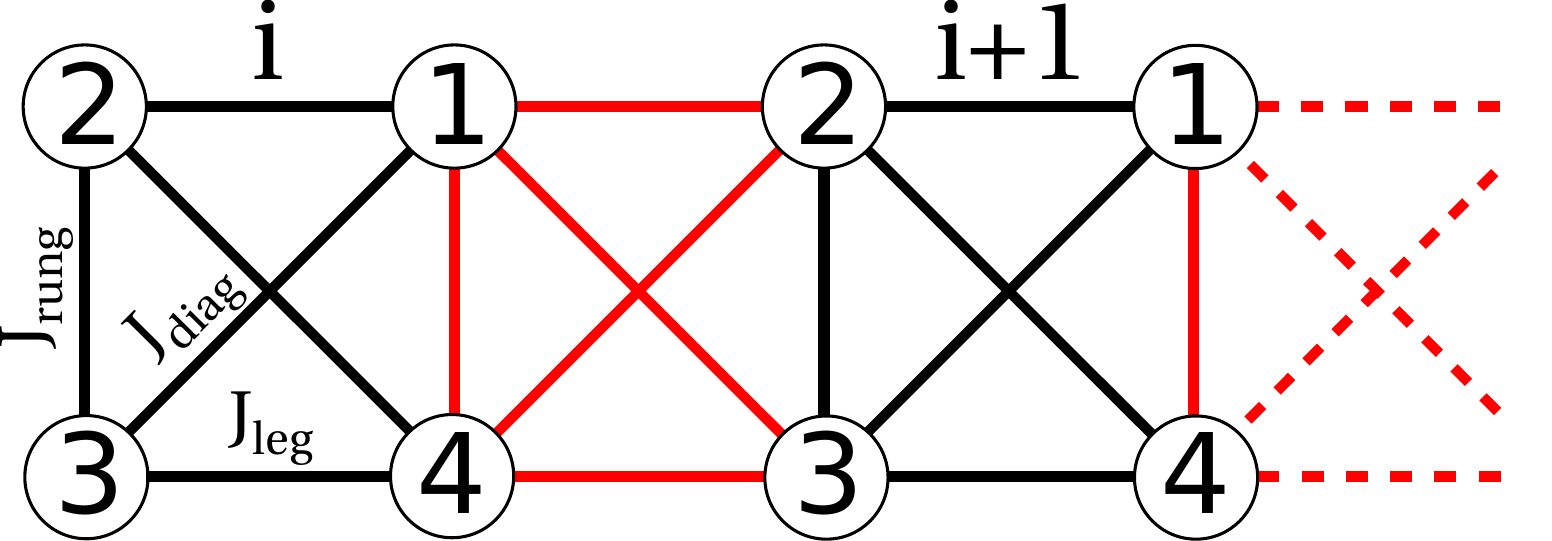}
\protect\protect\protect\caption{
Schematic view of the Hamiltonian of the 1D plaquette compass $Cx$--$Cz$
model, see Eq. (\ref{eq:Ham}). Black (red) lines stand for the XX (ZZ)
bonds connecting first and second neighbors. The system is naturally
divided into four-site cells (plaquettes) that are labeled by
$i=1,2,\dots,L$ and which carry internal site index $p=1,2,3,4$.}
\label{fig:plaqlad}
\end{figure}

We consider the Hamiltonian of the 1D plaquette compass $Cx$--$Cz$
model which can be written as follows,
\begin{eqnarray}
{\cal H} & = & \sum_{i=1}^{L}\left\{ J_{{\rm rung}}X_{i,2}X_{i,3}+J_{{\rm leg}}\left(X_{i,1}X_{i,2}+X_{i,3}X_{i,4}\right)\right\} \nonumber \\
& + & J_{{\rm diag}}\sum_{i=1}^{L}\left(X_{i,1}X_{i,3}+X_{i,2}X_{i,4}\right)
\nonumber \\
& + & \sum_{i=1}^{L}\left\{ J_{{\rm rung}}Z_{i,1}Z_{i,4}+J_{{\rm leg}}\left(Z_{i,1}Z_{i+1,2}+Z_{i,4}Z_{i+1,3}\right)\right\} \nonumber \\
& + & J_{{\rm diag}}\sum_{i=1}^{L}\left(Z_{i,1}Z_{i+1,3}+Z_{i,4}Z_{i+1,2}\right),
\label{eq:Ham}
\end{eqnarray}
where $X_{i,p}$ and $Z_{i,p}$ are the $\sigma^x$ and $\sigma^z$ Pauli
matrices for plaquette $i=1,\dots,L$ at site $p=1,2,3,4$, see Fig.
\ref{fig:plaqlad}. We consider all exchange interactions positive, i.e.,
antiferromagnetic (AF). A simpler $Cx$--$Cz$ model considered in Ref.
\cite{Brz14} can be recovered by setting $J_{{\rm rung}}=J_{{\rm leg}}$
and $J_{{\rm diag}}=0$. Here and below we
assume PBCs of the form $Z_{L+1,2}=Z_{1,2}$ and $Z_{L+1,3}=Z_{1,3}$.

There are two types of the symmetry operators specific to the model
(\ref{eq:Ham}), namely
\begin{eqnarray}
P_{i}^{z} & = & Z_{i,1}Z_{i,2}Z_{i,3}Z_{i,4},\nonumber \\
P_{i}^{x} & = & X_{i,1}X_{i,4}X_{i+1,2}X_{i+1,3}.
\end{eqnarray}
Following the spin transformations derived in Ref. \cite{Brz14} we
can find a block-diagonal form of the Hamiltonian ${\cal H}$ (\ref{eq:Ham})
in the common eigenspace of the symmetry operators. Here we will use
an alternative form of such transformation, as we consider it more
suitable to treat the present frustrated (generalized) problem, namely:
\begin{eqnarray}
X_{i,1} & = & r_{i}^{\star},\nonumber \\
X_{i,2} & = & r_{i}^{\star}s_{i}\tau_{i,2}^{x}\tau_{i,3}^{x},\nonumber \\
X_{i,3} & = & r_{i}^{\star}s_{i}\tau_{i,3}^{x},\nonumber \\
X_{i,4} & = & r_{i}^{\star}s_{i}\tau_{i+1,2}^{x},\label{eq:Xtrans}
\end{eqnarray}
and
\begin{eqnarray}
Z_{i,1} & = & s_{i}^{\star}\tau_{i,3}^{z},\nonumber \\
Z_{i,2} & = & s_{i-1}^{\star}\tau_{i,2}^{z},\nonumber \\
Z_{i,3} & = & s_{i-1}^{\star}r_{i}\tau_{i,2}^{z}\tau_{i,3}^{z},\nonumber \\
Z_{i,4} & = & s_{i}^{\star}.\label{eq:Ztrans}
\end{eqnarray}
Here $\tau_{i,2/3}^{x/z}$ are new Pauli operators at plaquette $i$
and sites $2$ and $3$ within the plaquette and $r_{i}$ and $s_{i}$
are the symmetry operators, namely
\begin{eqnarray}
r_{i}\equiv P_{i}^{z} & = & Z_{i,1}Z_{i,2}Z_{i,3}Z_{i,4},\nonumber \\
s_{i}\equiv P_{i}^{x} & = & X_{i,1}X_{i,4}X_{i+1,2}X_{i+1,3},\label{eq:rs}
\end{eqnarray}
and $r_{i}^{\star}$ , $s_{i}^{\star}$ are the anticommuting partners
of these operators that can be expressed in terms of initial Pauli
operators as,
\begin{eqnarray}
r_{i}^{\star} & = & X_{i,1},\nonumber \\
s_{i}^{\star} & = & Z_{i,4}.\label{eq:rs_star}
\end{eqnarray}
These operators will not be needed: $r_{i}^{\star}$ and $s_{i}^{\star}$
do not appear in the block-diagonal form of the Hamiltonian and $r_{i}$
and $s_{i}$ are good quantum numbers taking the values of $\pm1$.
The transformations of Eqs. (\ref{eq:Xtrans}) and (\ref{eq:Ztrans}),
whatever complicated they may be, are reversible. The inverse
transformations read:
\begin{eqnarray}
\tau_{i,2}^{x} & = & X_{i,2}X_{i,3},\nonumber \\
\tau_{i,3}^{x} & = & X_{i,3}X_{i,4}X_{i+1,2}X_{i+1,3},\label{eq:trinvX}
\end{eqnarray}
and
\begin{eqnarray}
\tau_{i,2}^{z} & = & Z_{i,2}Z_{i-1,4}\nonumber \\
\tau_{i,3}^{z} & = & Z_{i,1}Z_{i,4},\label{eq:trinvZ}
\end{eqnarray}
completed by the relations of Eqs. (\ref{eq:rs}) and (\ref{eq:rs_star}).
Note that however $r_{i}$ and $s_{i}$ can be alternatively called
$r_{i}\equiv\tau_{i,1}^{z}$ and $s_{i}\equiv\tau_{i,4}^{x}$, as they
commute with the pseudospins $\tau_{i,2}^{x/z}$ and $\tau_{i,3}^{x/z}$.
This \textit{does not} automatically imply that $r_{i}^{\star}$ and
$s_{i}^{\star}$ can be indeed identified with $\tau_{i,1}^x$ and
$\tau_{i,4}^z$, we find that $r_i^{\star}$ anticommutes with
$\tau_{i,3}^{z}$ and $s_i^{\star}$ with $\tau_{i,3}^{x}$. The block
diagonal Hamiltonian has the following form:
\begin{eqnarray}
{\cal H} & = & \sum_{i=1}^{L}\left\{ J_{{\rm rung}}\tau_{i,2}^{x}+J_{{\rm leg}}
\left(s_{i}\tau_{i,2}^{x}\tau_{i,3}^{x}+\tau_{i,3}^{x}\tau_{i+1,2}^{x}\right)\right\}
\nonumber \\
& + & J_{{\rm diag}}\sum_{i=1}^{L}
\left(s_{i}+\tau_{i,2}^{x}\tau_{i+1,2}^{x}\right)\tau_{i,3}^{x}\nonumber \\
& + & \sum_{i=1}^{L}\left\{ J_{{\rm rung}}\tau_{i,3}^{z}+J_{{\rm leg}}
\left(r_{i}\tau_{i,2}^{z}\tau_{i,3}^{z}+\tau_{i,3}^{z}\tau_{i+1,2}^{z}\right)\right\}
\nonumber \\
& + & J_{{\rm diag}}\sum_{i=1}^{L}
\left(1+r_{i}\tau_{i-1,3}^{z}\tau_{i,3}^{z}\right)\tau_{i,2}^{z}.
\label{eq:Ham2}
\end{eqnarray}
From the computational point of view it is worth to notice that thanks
to the symmetries we reduce the number of quantum degrees of freedom by
half, i.e., instead of four initial Pauli operators
$\left\{ X_{i,1},X_{i,2},X_{i,3},X_{i,4}\right\}$ and their anticommuting
partners per each plaquette (see Fig. \ref{fig:plaqlad}), we now have only
two of them, $\left\{ \tau_{i,2}^{z},\tau_{i,3}^{z}\right\}$, with
anticommuting partners and two good quantum numbers $\left\{r_i,s_i\right\}$
taking values $\pm1$. Thus instead of plaquettes with internal site index
$p=1,2,3,4$ we now deal with dimers with $p=2,3$. From the fundamental
point of view the specific pattern of the $\left\{ r_{i},s_{i}\right\}$
Ising spins is an important characterization of model's ground state
at given values of the interactions, as we shall see further on.

Note that in the absence of diagonal interaction, i.e., at
$J_{{\rm diag}}=0$, the model is equivalent to the 1D XY model in the
external XY field. To see it clearly one can redefine the Pauli
operators to get rid of $p=2,3$ indices and gauge away the $r_{i}$ and
$s_i$ phase from the interaction term (strictly speaking this can be done
completely only for an open chain so we neglect here the closing
$\langle L,1\rangle$ bond). To do this we use a following transformation:
\begin{eqnarray}
\tau_{i,2}^{x} & = & s_{1}s_{2}s_{3}\dots s_{i-1}\tau'{}_{2i-1}^{x},\nonumber \\
\tau_{i,3}^{x} & = & s_{1}s_{2}s_{3}\dots s_{i}\tau'{}_{2i}^{x},
\end{eqnarray}
and
\begin{eqnarray}
\tau_{i,2}^{z} & = & r_{1}r_{2}r_{3}\dots r_{i-1}\tau'{}_{2i-1}^{z},\nonumber \\
\tau_{i,3}^{z} & = & r_{1}r_{2}r_{3}\dots r_{i}\tau'{}_{2i}^{z},
\end{eqnarray}
to get
\begin{eqnarray}
{\cal H}'_{{\rm OBC}}\!\left(J_{{\rm diag}}\!=\!0\right) & \!=\!
& J_{{\rm rung}}\!\sum_{i=1}^{L}\left(R_{i}\tau'{}_{2i}^{z}\!
+\! S_{i-1}\tau'{}_{2i-1}^{x}\right)\nonumber \\
& \!+\! & J_{{\rm leg}}\!\sum_{i=1}^{2L-1}\!
\left(\tau'{}_{i}^{z}\tau'{}_{i+1}^{z}\!
+\!\tau'{}_{i}^{x}\tau'{}_{i+1}^{x}\right)\!,
\end{eqnarray}
with
\begin{eqnarray}
R_{i} & \equiv & r_{1}r_{2}r_{3}\dots r_{i}.\nonumber \\
S_{i-1} & \equiv & s_{1}s_{2}s_{3}\dots s_{i-1}.
\end{eqnarray}
Here the chain has open boundary conditions (OBC). Clearly the interaction
part is the XY model that can be solved by a Jordan-Wigner transformation
to get a solution in terms of free fermions with cosine dispersion.
On the other hand, the linear part makes the model unsolvable as it
becomes nonlocal and of infinite order in the thermodynamic limit
after the transformation. Nevertheless, we find the $Cx$--$Cz$ model
written in this form simpler than in the original form introduced
in Ref. \cite{Brz14} --- see the Appendix \ref{sec:appA} for the
transformation linking these two forms.

Another interesting limit is when $J_{{\rm rung}}=0$ and either
$J_{{\rm leg}}=0$ or $J_{{\rm diag}}=0$.
Looking at Fig. \ref{fig:plaqlad} one can easily see that in this
limit the system splits into two independent subsystems and each of
them is described by the so-called 1D compass model \cite{Brz07},
with XX and ZZ interactions alternating on even/odd bonds. Away from
these points the two 1D compass models start to interact in a very
complex way.

\section{High symmetry line}
\label{sec:line}

Apart from the local parities which are the symmetries of the model
for any choice of its exchange parameters $J$'s, there is a special
line in the parameter space, namely
\begin{equation}
J_{{\rm leg}}=J_{{\rm diag}},
\label{line}
\end{equation}
where the model has extra symmetries. These are interchanges of the two
spins located at every rung of the ladder in Fig. \ref{fig:plaqlad} ---
it is easy to notice that such an operation done on a single rung will
interchange leg and diagonal bonds within the two plaquettes adjacent
to this rung. When Eq. (\ref{line}) is obeyed, this has no effect on
Hamiltonian (\ref{eq:Ham}). Such interchange for a $(2,3)$ rung can be
realized by a spin-interchange operator known from the so-called Kumar
model \cite{Kum13}, i.e.,
\begin{equation}
{\cal X}_{i}^{2,3}=\frac{1}{2}\left(
1+\vec{\sigma}_{i,2}\cdot\vec{\sigma}_{i,3}\right),
\end{equation}
where $\vec{\sigma}_{i,p}=\{X_{i,p},Y_{i,p},Z_{i,p}\}$. Similarly one
can define the spin interchange operator ${\cal X}_{i}^{1,4}$ for rungs
$(1,4)$.

If Eq. (\ref{line}) applies, the Hamiltonian commutes with
${\cal X}_{i}^{2,3}$ and ${\cal X}_{i}^{1,4}$ for every $i$,
whose spectrum consists of one eigenvalue $\lambda_{s}=-1$ for a spin
singlet on a rung and three eigenvalues $\lambda_{t}=1$ for a spin
triplet. This knowledge can be used to rewrite the Hamiltonian of Eq.
(\ref{eq:Ham}) at constraint Eq. (\ref{line}) in the form of
${\cal H}_{{\rm sym}}\equiv{\cal H}(J_{{\rm leg}}=J_{{\rm diag}})$
given by a following equation,
\begin{eqnarray}
{\cal H}_{{\rm sym}} & = & 2J_{{\rm rung}}\sum_{i=1}^{L}\left\{ \left(S_{i,2}^{x}\right)^{2}+\left(S_{i,1}^{z}\right)^{2}-1\right\} \nonumber \\
 & + & 4J_{\rm leg}\sum_{i=1}^{L}
 \left(S_{i,2}^{x}S_{i,1}^{x}+S_{i,1}^{z}S_{i+1,2}^{z}\right),
 \label{eq:Hsym}
\end{eqnarray}
where $S_{i,3/4}^{x/z}$ are the spin $S=1$ or $S=0$ operators being
the sums of two $S=1/2$ spin operators on every rung, i.e.,
\begin{eqnarray}
S_{i,2}^{x}=\frac{1}{2}\left(X_{i,2}+X_{i,3}\right) & ,\quad & S_{i,2}^{z}=\frac{1}{2}\left(Z_{i,2}+Z_{i,3}\right),\nonumber \\
S_{i,1}^{x}=\frac{1}{2}\left(X_{i,1}+X_{i,4}\right) & ,\quad & S_{i,1}^{z}=\frac{1}{2}\left(Z_{i,1}+Z_{i,4}\right).
\end{eqnarray}
The spin-interchange symmetry present in ${\cal H}_{{\rm sym}}$
guarantees that the total spin $S$ on every site $(i,2)$ and $(i,1)$ is
a good quantum number. Note that in analogy to the previous Section
we mapped a ladder Hamiltonian onto a model of dimerized chain but
here the building blocks are $S=1$ or $S=0$ spins. This new Hamiltonian
still has the parity symmetries which are now given by
\begin{eqnarray}
P_{i}^{z} & = & \left(2\left(S_{i,2}^{z}\right)^{2}-1\right)
\left(2\left(S_{i,1}^{z}\right)^{2}-1\right),\\
P_{i}^{x} & = & \left(2\left(S_{i,1}^{x}\right)^{2}-1\right)
\left(2\left(S_{i+1,2}^{x}\right)^{2}-1\right),
\end{eqnarray}
whose meaning is now the parity of number of $S_i^z=0$ or $S_i^x=0$
eigenvalues on every $(2,1)$ or $(1,2)$ bond respectively.

Note that
${\cal H}_{{\rm sym}}$ has a degenerate manifold of very simple
eigenstates with energy $E_{0}=-2J_{rung}L$ and degeneracy $d=2^{2L}$.
These states we construct by putting on every site $(i,2)$ either
total spin $S=0$ or spin $S=1$ with $S_{i,2}^{x}=0$ and on every
site $(i,1)$ either total spin $S=0$ or spin $S=1$ with $S_{i,1}^z=0$.
For the original ladder this means that on every rung $(2,3)$ we
put a spin singlet or a spin triplet with zero projection on $x$
spin axis and on every rung $(1,4)$ we put a spin singlet or a spin
triplet with zero projection on $z$ spin axis. Equivalently one can
take a symmetric or antisymmetric combination of singlet and triplet
and then for every rung $(2,3)$ we can choose between states
$\left|\leftarrow\rightarrow\right\rangle$
and $\left|\rightarrow\leftarrow\right\rangle$ and for every rung
$(1,4)$ we can have either $\left|\uparrow\downarrow\right\rangle $
or $\left|\downarrow\uparrow\right\rangle $. In this way we can produce
$2^{2L}$ rung-product states with energy $E_{0}$. Such states belong to
the ground state manifold of the model with the constraint Eq.
(\ref{line}) and large enough $J_{{\rm rung}}$, and we will call them
nematic in the following Sections.

Finally, we note that by setting $J_{{\rm rung}}=0$ in Eq.
(\ref{eq:Hsym}), we get a Hamiltonian whose low-energy sector realizes
a 1D compass model with spins $S=1$. Combining Eqs. (\ref{eq:Ham2}) and
(\ref{eq:Hsym}) at $J_{{\rm rung}}=0$ we see that the present plaquette
model realizes a crossover within a class of 1D compass models ---
setting $J_{{\rm leg}}=1$ and increasing $J_{{\rm diag}}$ from 0 to 1
changes spins $S=1/2$ to spins $S=1$. We note that the $S=1$ 1D compass
model was studied before \cite{You13} and it was shown that for such a
choice of parameter values as here the ground state of the model is
unique, disordered and gapped. We also found that it is strongly
dimerized, i.e., looking at two-point correlation functions only the
correlations that enter the Hamiltonian (\ref{eq:Hsym}) are non-vanishing.

\section{Possible quantum phases}
\label{sec:phd}

\subsection{A single plaquette with frustration}
\label{sec:box}

The ground state of the model given by the block-diagonal Hamiltonian
of Eq. (\ref{eq:Ham2}) can occur in different subspaces of the symmetry
operators (i.e., different diagonal blocks) labeled by the quantum
numbers $\left\{ r_{1},\dots,r_{L},s_{1},\dots,s_{L}\right\} $,
depending on the values of exchange parameters,
$\{J_{{\rm rung}},J_{{\rm leg}},J_{{\rm diag}}\}$. This is a more
complex situation than in the case of unfrustrated $Cx$--$Cz$ model
where the ground state was found always in the subspace
$r_{i}\equiv s_{i}\equiv1$ \cite{Brz14}. The difference is a
manifestation of the intrinsic frustration induced by the diagonal bonds.

Let us consider first a single (open) $Z$-plaquette shown in Fig.
\ref{fig:plaqlad}, described by a $4$-site Hamiltonian of the form,
\begin{eqnarray}
{\cal H}_{\Box} & = & J_{{\rm rung}}Z_{2}Z_{3}
+J_{{\rm leg}}\left(Z_{1}Z_{2}+Z_{3}Z_{4}\right)\nonumber \\
 & + & J_{{\rm diag}}\left(Z_{1}Z_{3}+Z_{2}Z_{4}\right),\label{box}
\end{eqnarray}
where pseudospins at sites $i=1$ and $i=4$ do not interact as this
bond belongs to the next plaquette, see Fig. \ref{fig:plaqlad}. We
set a constraint in the parameter space,
\begin{equation}
J_{{\rm diag}}=2-J_{{\rm leg}},\label{unit}
\end{equation}
and we change $J_{{\rm leg}}$ in the interval $0\leq J_{{\rm leg}}\leq2$
to interpolate between the two unfrustrated $Cx$--$Cz$ models. Equation
(\ref{unit}) serves also to determine the units of dimensionless
exchange parameters. Plaquette frustration vanishes when
$J_{{\rm diag}}=0$, or in the equivalent limit when $J_{{\rm leg}}=0$
(but $J_{{\rm diag}}=2$) --- in both limits the plaquette spins form an
open chain and one recovers a unit of the unfrustrated $Cx$--$Cz$ model
of Ref. \cite{Brz14}. All eigenstates are classical and have degeneracy
$d=2$ (with equivalent configurations obtained by reversing all spins).

One can easily check that three distinct configurations exist which
become ground states in various regions of parameters: \hfill{}\\
(i) In the area of
$J_{{\rm rung}}\leq2\left(1-\left|1-J_{{\rm leg}}\right|\right)$, the
pseudospin configuration in the ground state is the one that satisfies
leg and diagonal bonds but the rung bond is frustrated. This state is,
\begin{eqnarray}
\left|\psi_{0}\right\rangle =\left|\uparrow\right\rangle _{1}
\left|\downarrow\right\rangle _{2}\left|\downarrow\right\rangle _{3}
\left|\uparrow\right\rangle _{4}.\label{psi1}
\end{eqnarray}
(ii) For stronger
$J_{{\rm rung}}>2\left(1-\left|1-J_{{\rm leg}}\right|\right)$ and
$J_{{\rm leg}}<1$, one finds the ground state that optimizes only rung
and diagonal bonds, namely
\begin{eqnarray}
\left|\psi_{-1}\right\rangle =\left|\downarrow\right\rangle _{1}
\left|\downarrow\right\rangle _{2}\left|\uparrow\right\rangle _{3}
\left|\uparrow\right\rangle _{4}.\label{psi2}
\end{eqnarray}
(iii) Finally, for
$J_{{\rm rung}}>2\left(1-\left|1-J_{{\rm leg}}\right|\right)$ and
$J_{{\rm leg}}>1$ we find the ground state that optimizes only rung
and leg bonds which is
\begin{eqnarray}
\left|\psi_{1}\right\rangle =\left|\uparrow\right\rangle _{1}
\left|\downarrow\right\rangle _{2}\left|\uparrow\right\rangle _{3}
\left|\downarrow\right\rangle _{4}.\label{psi3}
\end{eqnarray}
The states (\ref{psi2}) and (\ref{psi3}) are degenerate along the
line $J_{{\rm leg}}=1$ for $J_{{\rm rung}}>2$ in Fig. \ref{fig:PD4}.
The lines between the $|\psi_{0}\rangle$ and $|\psi_{1}\rangle$
($|\psi_{-1}\rangle$) states set the stage and different phases of
a ladder consisting of $Z$ and $X$ plaquettes presented in Fig.
\ref{fig:plaqlad} develop around them, as we show below. We can expect
that in the quantum many-body regime the intracell frustration discussed
here will result in very non-trivial configurations.

\subsection{Phase diagram for a ladder of $L=4$ plaquettes}
\label{sec:PD4}

\begin{figure}[b!]
\includegraphics[width=1\columnwidth]{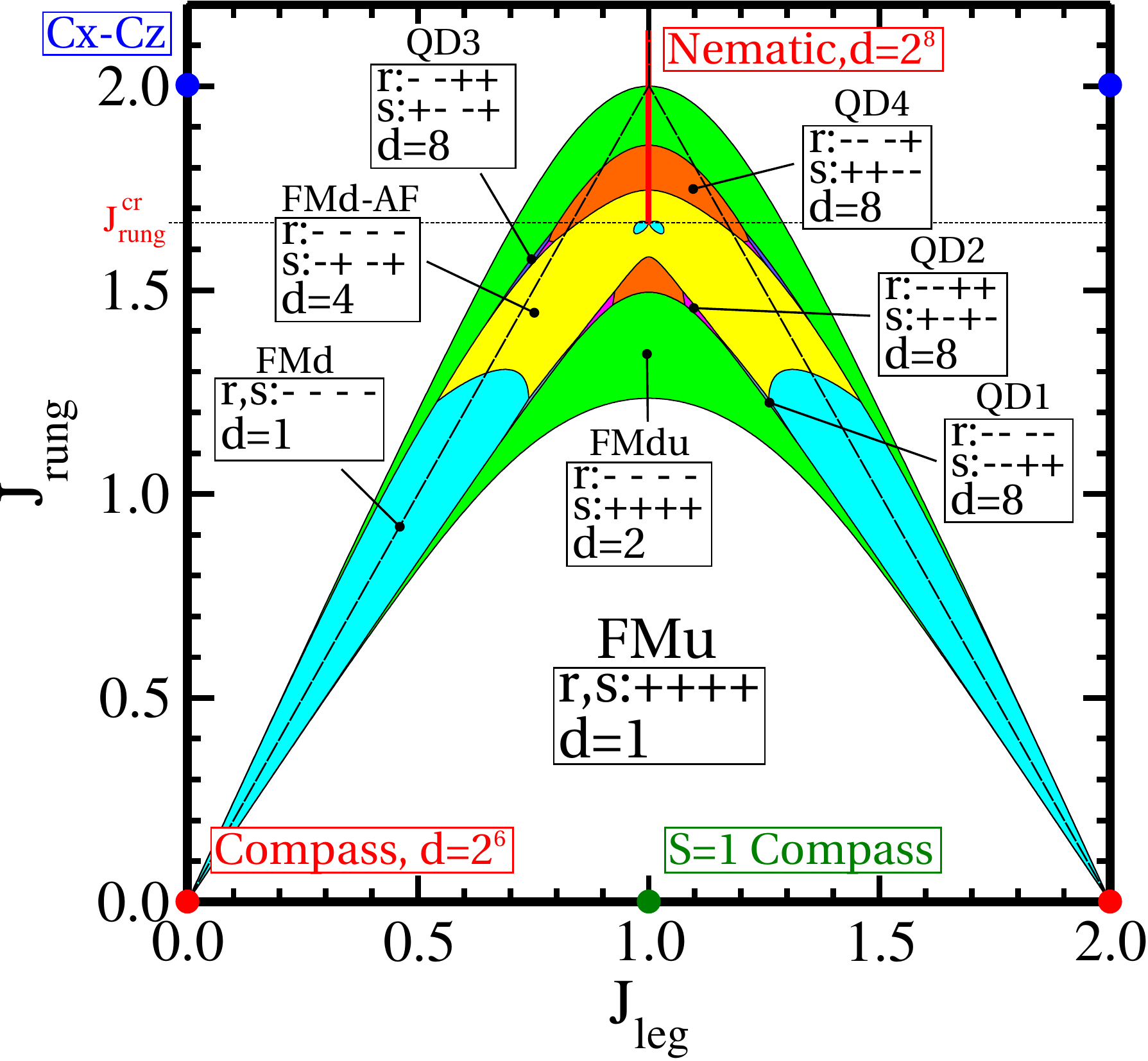}
\protect\protect\protect\caption{
Ground state phase diagram for a $L=4$ plaquette ladder with PBCs
as function of $J_{{\rm leg}}$ and $J_{{\rm rung}}$. The exchange on
diagonal bonds $J_{{\rm diag}}$ is set following Eq. (\ref{unit}).
The ground states
(phases) are shaded in color or white and labeled according to their
symmetry patterns $\left\{ r_{1},\dots,r_4;s_{1},\dots,s_4\right\}$,
which are indicated together with their degeneracies $d$. Red dots
are the 1D compass points, blue dots indicate the simple $Cx$--$Cz$
states and green dot is $S=1$ 1D compass point. The critical value
$J_{{\rm rung}}^{{\rm cr}}$ for a nematic state is indicated by
horizontal line at $J_{{\rm leg}}=1.0$. The boundaries separating
classical states $\{\psi_{0},\psi_{-1},\psi_{1}\}$ obtained for a
single frustrated $Z$-plaquette (\ref{box}) are indicated by thin
dashed lines. \label{fig:PD4}}
\end{figure}

\begin{figure*}[t!]
\includegraphics[width=1\textwidth]{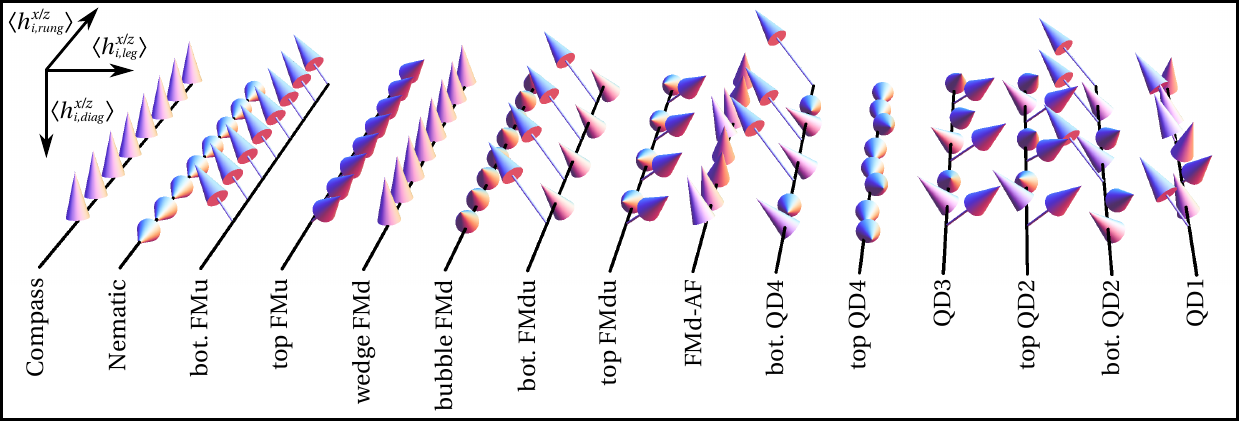}
\protect\protect\protect\caption{
Visualization of the configurations which characterize different
types of order in the phase diagram of Fig. \ref{fig:PD4}. The solid
lines indicate both the system itself and different phases or different
domains of one phase which are labeled below. Arrows show the strength
of the XX or ZZ interactions within consecutive plaquettes of the
ladder such that their 3D components correspond to the ground-state
average of the rung, leg and diagonal interactions denoted as
$\langle h_{i,{\rm rung}}^{x/z}\rangle$,
$\langle h_{i,{\rm leg}}^{x/z}\rangle$, and
$\langle h_{i,{\rm diag}}^{x/z}\rangle$, respectively. The reference
frame is given in the left top corner. Arrows along the solid lines
mean that only rung interactions are non-zero, vertical arrows mean
diagonal interactions, and horizontal ones on the legs only.}
\label{fig:cnf4}
\end{figure*}

Indeed, in Fig. \ref{fig:PD4} we find a complex phase diagram found
for a system of $L=4$ plaquettes with PBCs, containing $10$ different
pseudospin configurations. All the phase boundaries mark the level
crossings between the ground states from the different invariant
subspaces of the Hamiltonian (\ref{eq:Ham}) labeled by the quantum
numbers $\{r_i,s_i\}$. The phase diagrams were obtained in the
following way: first the phase space was discretized by a sufficiently
dense lattice of points (typically $100\times100$ points for
$0\le J_{{\rm leg}}\le1$ and $0\le J_{{\rm leg}}\le2.2$) and for every
point all the symmetry-nonequivalent subspaces were searched for the
ground state to found the one with the lowest energy. Then, knowing
what the phases are and how approximately they are located, the
bisection algorithm was used to establish smooth boundaries between
them. The dominant one was found before in the unfrustrated $Cx$--$Cz$
model, where $r_{i}\equiv s_{i}\equiv1$. This phase we call FMu because
$\{r_{i}\}$ and $\{s_{i}\}$ variables can be seen as Ising spins that
are here in the FM configuration, all pointing up. This phase consists
of two parts, the bottom and the top one. These parts are separated by
a window of other configurations
that stem from the compass points located in the bottom corners of
the phase diagram. This window contains the classical phase boundaries
discussed for a single plaquette in Sec. \ref{sec:box} and ends exactly
at its critical point $(J_{{\rm leg}},J_{{\rm rung}})=(1.0,2.0)$.

Firstly, we note that the diagram is symmetric around the high-symmetry
line $J_{{\rm leg}}=1$ starting from the $S=1$ compass point at
$J_{{\rm rung}}=0$, which is deeply embedded in the FMu phase. Going
up from this point we increase the values of the quadrupolar external
fields in the high-symmetry Hamiltonian (\ref{eq:Hsym}) and in the
end we leave the FMu phase. Note that the ground state remains gapped
along the whole $J_{{\rm leg}}=1$ line as it was at $J_{{\rm rung}}=0$.
Moving left or right from this line we lower the symmetry of the problem
- the $S=1$ or $S=0$ spins placed on every rung of the ladder dissociate
into pairs of spins $S=1/2$ up to extreme case of compass points
with exponential degeneracy at $J_{{\rm leg}}=0$ and $J_{{\rm rung}}=0$.
This 'quenching' of pairs of spins $S=1/2$ is similar to the formation
of $J=0$ singlets out of spins $S=1/2$ and orbital moments $L=1/2$
in presence of strong spin-orbit coupling \cite{Jac09}.

In the narrow shoulders of the window starting from $J_{{\rm rung}}=0$
and either $J_{{\rm leg}}=0$ or $J_{{\rm leg}}=2.0$ we find two
wedges of the FMd phase being exactly opposite to the FMu configuration,
where all the symmetries have $-1$ eigenvalues. This phase is stable
for not too large $J_{{\rm rung}}$ and away from the high symmetry
point $J_{{\rm leg}}=J_{{\rm diag}}=1.0$. Quite surprisingly tiny
bubbles of this phase can be found at the onset of the phase that
we call nematic, exactly at $J_{{\rm leg}}=J_{{\rm diag}}=1.0$ and
$J_{{\rm rung}}=J_{{\rm rung}}^{\rm cr}\simeq1.664$. Centrally around
$J_{{\rm leg}}=1$ at the very top and bottom of the window we find
a phase which is partially FMu and FMd, where $r_{i}\equiv-1$ and
$s_{i}\equiv1$, namely the FMdu phase. This configuration is doubly
degenerate because here is is permitted to uniformly interchange $r_{i}$
and $s_{i}$ quantum numbers and set $r_{i}\equiv1$ and $s_{i}\equiv-1$.

More generally, the main source of degeneracies found in the phase
diagram of Fig. \ref{fig:PD4} is the invariance of the block diagonal
Hamiltonian of Eq. (\ref{eq:Ham2}) with respect to translations and
to $\{r_{i}\}\leftrightarrow\{s_{i}\}$ interchange. Because of this
we are always allowed to cyclically translate the symmetry quantum
numbers, i.e., going from the subspace
$\left\{ r_{1},\dots,r_{L};s_{1},\dots,s_{L}\right\}$ to
$\left\{ r_{L},r_{1}\dots,r_{L-1};s_{L},s_{1},\dots,s_{L-1}\right\}$
will not affect the energies. Similarly, we are allowed to translate
the initial ladder of Fig. \ref{fig:plaqlad} by a one lattice spacing
along the ladder legs provided that we also interchange all $X$
operators with the $Z$ ones. Thus going from the subspace
$\left\{ r_{1},\dots,r_{L};s_{1},\dots,s_{L}\right\}$
to $\left\{ s_{L},s_{1},s_{2}\dots,s_{L-1};r_{1},\dots,r_{L}\right\}$
will not affect the energies as long as the $X$- and $Z$-bonds are
equivalent. For simplicity this operation on the
$\left\{ r_{i},s_{i}\right\}$ subspace labels will be called the
$r\leftrightarrow s$ interchange, keeping in mind that there is also a
translation of the $s_{i}$ quantum numbers involved. Note that a very
similar discussion concerns the analogous $\{r_{i},s_{i}\}$ quantum
numbers in the 2D compass model, see Refs. \cite{Brz10}.

Going further towards the center of the window in the phase diagram
of Fig. \ref{fig:PD4} we find the last ordered phase which is the
FMd-AF one. In this phase the $\{r_{i}\}$ quantum numbers are all equal
to $-1$ as in FMd phase but the $s_{i}$ ones are alternating as in the
AF state, hence we use the label AF. This phase has a degeneracy
of $d=8$ coming both from translation and the $r\leftrightarrow s$
interchange discussed above. In the similar area of the phase diagram
we find a phase with no particular order in the eigenvalues of the
symmetries which we call QD4, meaning quantum disorder. There are
three other phases of this type, QD1-QD3, but all of them are stable
only in tiny regions of the phase diagram, i.e., close to quantum
critical points where three different phases meet.

Finally, we find special phases of measure zero in the phase diagram
which have macroscopic degeneracy that cannot be explained only by
the translation and the $r\leftrightarrow s$ interchange symmetry.
These are the already mentioned compass points where we recover the
degeneracy of two independent 1D compass models
$d=\left(2^{L-1}\right)^2$ \cite{Brz07}, and less expected nematic
phase with degeneracy $d=2^{2L}$. In the latter phase the degeneracy
comes from the fact that both leg and diagonal interactions in this
phase have zero expectation value and only the rungs give finite
contribution to the ground state energy. Thus according to Eq.
(\ref{eq:Ham2}) the quantum numbers $\{r_{i}\}$ and $\{s_{i}\}$ do not
affect the energy and can be arbitrary. Quite remarkably the onset of
the nematic phase does not coincide with the classical value of
$J_{{\rm rung}}=2.0$ but is placed much lower at
$J_{{\rm rung}}^{{\rm cr}}\simeq1.664$. This is clearly the effect of
frustration which is not included in a single $Z$-plaquette, see Sec.
\ref{sec:box} and comes from the competition between the XX and ZZ
bonds that want to order pseudospins along two perpendicular directions.

\begin{figure}[t!]
\includegraphics[width=1\columnwidth]{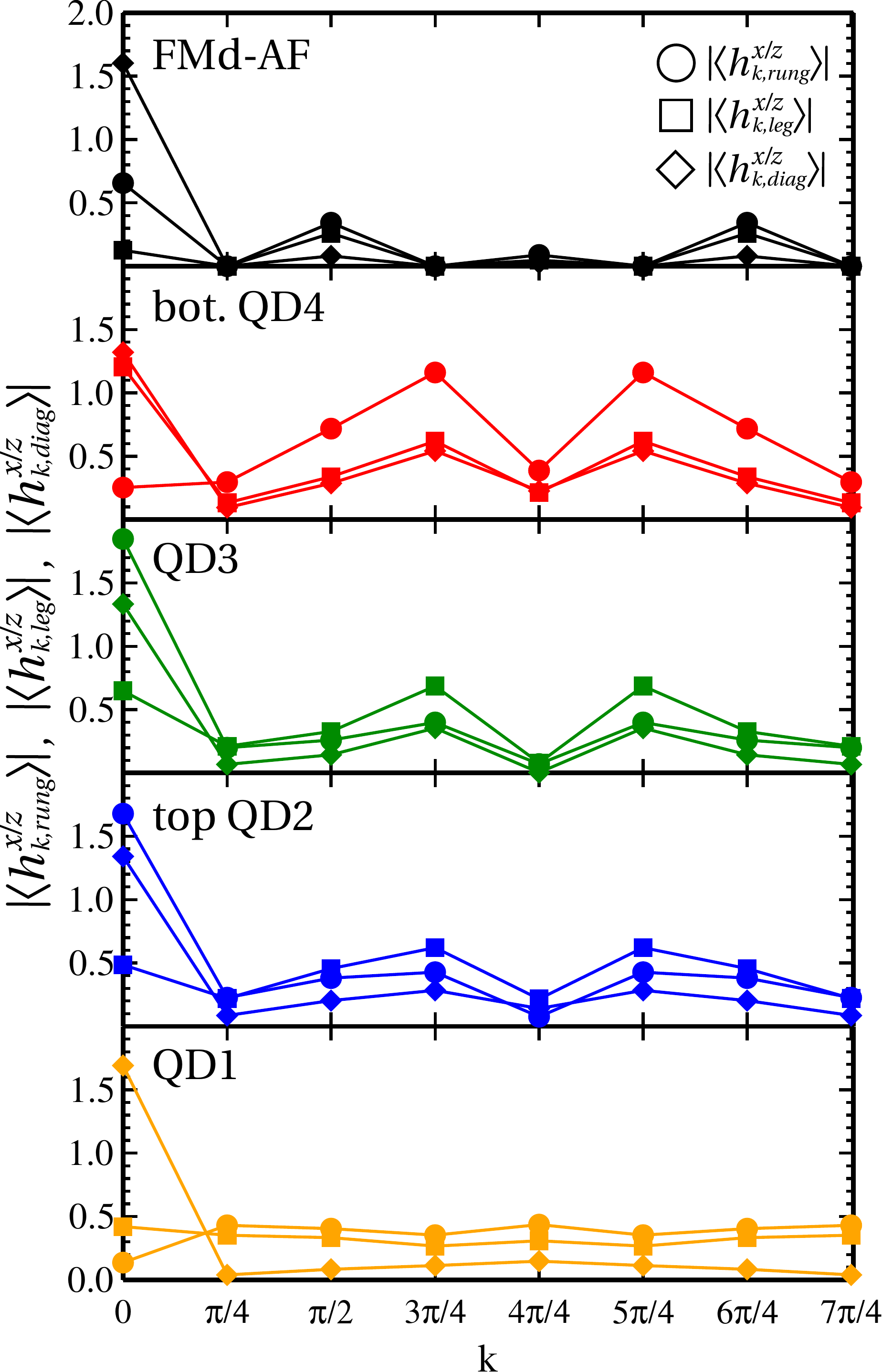}
\protect\protect\caption{
Absolute values of the Fourier transforms of the bond averages $
\langle h_{i,{\rm rung}}^{x/z}\rangle$,
$\langle h_{i,{\rm leg}}^{x/z}\rangle$, and
$\langle h_{i,{\rm diag}}^{x/z}\rangle$, shown in Fig. \ref{fig:cnf4}
in the phases FMd-AF, bot. QD4, QD3, top QD2 and QD1. Different symbols
--- circles, squares, diamonds correspond
to rung, leg and diagonal bonds respectively. \label{fig:struct_4}}
\end{figure}

The different ground state configurations (phases) realized by the
model can be conveniently characterized by a 3D vector field living
on one leg of the initial ladder. In three components of such vectors
we can encode the average values of three different bonds outgoing
from the point at which the vector is anchored --- these are leg,
rung and diagonal bonds. Because the two legs of the ladder are perfectly
equivalent, it is enough to check the following bond operators labeled
by a site index $j$ along a single leg, $1\leq j\leq2L$,
\[
h_{j,{\rm rung}}^{x/z}=\begin{cases}
X_{i,2}X_{i,3}\equiv\tau_{i,2}^{x} & j=(2i-1)\\
Z_{i,1}Z_{i,4}\equiv\tau_{i,3}^{z} & j=2i
\end{cases},
\]
\[
h_{j,{\rm leg}}^{x/z}=\begin{cases}
X_{i,1}X_{i,2}\equiv s_{i}\tau_{i,2}^{x}\tau_{i,3}^{x} & j=(2i-1)\\
Z_{i,1}Z_{i+1,2}\equiv\tau_{i,3}^{z}\tau_{i+1,2}^{z} & j=2i
\end{cases},
\]
\[
h_{j,{\rm diag}}^{x/z}=\begin{cases}
X_{i,1}X_{i,3}\equiv s_{i}\tau_{i,3}^{x} & j=(2i-1)\\
Z_{i,1}Z_{i+1,3}\equiv\\
r_{i+1}\tau_{i,3}^{z}\tau_{i+1,2}^{z}\tau_{i+1,3}^{z} & j=2i
\end{cases},
\]
where $i$ labels the plaquettes, $1\leq i\leq L$. Here we keep in
mind that these bonds are of the XX(ZZ) type for odd(even) points
along the ladder's leg. In this way such vector fields tell us all
about the NN interactions in a given phase. In this context we can
consider a periodicity of a given phase. If the XX and ZZ interactions
are perfectly balanced then we can expect that for simplest ground
states the bond averages will be the same for every point along the
leg and consequently all the vectors will be the same. Such configuration
respects the translational and the $r\leftrightarrow s$ interchange
symmetry of the initial ladder, as one could expect from the ground
state.

To understand better the physical meaning of phases found in the phase
diagram of Fig. \ref{fig:PD4} we present the ground-state values of all
different bonds of the initial Hamiltonian shown in Fig.
\ref{fig:plaqlad}. In Fig. \ref{fig:cnf4} we show a 3D representation
of the ground-state averages of the above operators as functions of the
position $i$ at 15 representative points in the phase diagram. These
points are placed in all different phases but also in two pieces of one
phase; for instance we have bottom and top part of the FMu phase
meaning simply the low- and high-$J_{{\rm rung}}$ parts visible in the
phase diagram. This terminology we also apply to few other phases and
concerning the FMd phase we write 'wedge' and 'bubble' to distinguish
between the tiny piece of FMd close to $J_{{\rm leg}}=1.0$ and
$J_{{\rm rung}}=J_{{\rm rung}}^{\rm cr}$ and the rest of it.

Figure \ref{fig:PD4} first shows particular limits, like that:
(i) in the nematic phase only the rung bonds are non-zero, or
(ii) one has only diagonals or legs non-zero at the compass points, or
(iii) the bottom FMu differs from the top FMu phase by the polarization
of the rung bonds; in the bottom part the rungs can be neglected but
in the top part they are satisfied.
Further on, we can see that the
wedge part of the FMd phase looks very similar to the compass point
from which it stems and in the bubble FMd phase the rung bonds are
much more favored. The combined FMdu phase exhibits a strong period
$2$ alternation of the bond values whereas in the FMd-AF one the
period is $4$. Thus we say that the FMdu phase is dimerized in the
sense that going along the ladder we will observe stronger/weaker
alternation of a given type of a bond, i.e., leg, rung or diagonal one.
Analogously, the FMd-AF is tetramerized. Finally, we find why the QD
phases are really disordered as they exhibit no periodicity. This means
that translational invariance is completely broken in these phases.
Of course, if we average over the degenerate $\left\{r_i,s_i\right\}$
configurations then the translational invariance will be recovered
but the system will not gain any energy by forming such a superposition.

Finally, in Fig. \ref{fig:struct_4} we show the Fourier transforms
of the bond averages $\langle h_{i,{\rm rung}}^{x/z}\rangle$,
$\langle h_{i,{\rm leg}}^{x/z}\rangle$, and
$\langle h_{i,{\rm diag}}^{x/z}\rangle$ in the phases with longest
periods, namely FMd-AF, bot. QD4, QD3, top QD2 and QD1. These are
defined as,
\begin{equation}
\langle h_{k,{\rm bond}}^{x/z}\rangle\equiv\frac{1}{\sqrt{2L}}
\sum_{j=1}^{2L}e^{ikj}\langle h_{j,{\rm bond}}^{x/z}\rangle,
\label{eq:struct}
\end{equation}
with ${\rm bond}={\rm rung},{\rm leg},{\rm diag}$, $k=\frac{2\pi}{2L}n$
and $n=0,1,\dots,2L-1$. In all phases shown in Fig. \ref{fig:struct_4}
we see a dominant ferromagnetic order at least for one type of bond
meaning large $k=0$ Fourier component. In case of FMd-AF phase we
have four non-vanishing Fourier components for any type of bond which
indicates a four-site unit cell order. In case of the QD phases all
the components are non-vanishing indicating no translational symmetry.

\begin{figure}[t!]
\includegraphics[width=1\columnwidth]{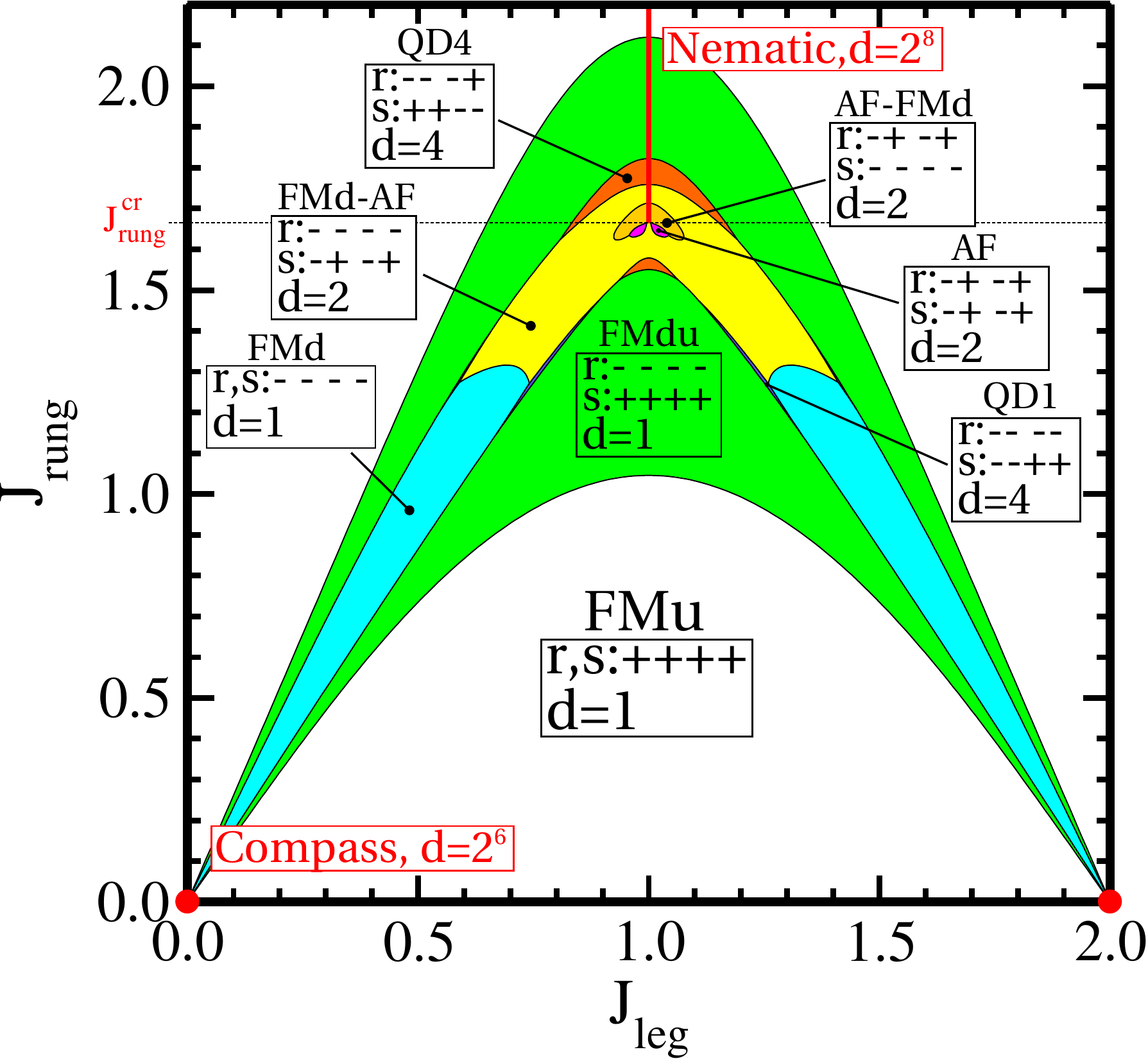}
\protect\protect\caption{
Ground state phase diagram for a system with $L=4$ plaquettes and
all $Z$-bonds rescaled by a factor of $\gamma=0.9$ as function of
$J_{{\rm leg}}$ and $J_{{\rm rung}}$. The exchange on diagonal bonds
$J_{{\rm diag}}$ is set following  Eq. (\ref{unit}). The ground states
are shaded in color or white and labeled according to their symmetry
patterns, $\left\{ r_{1},\dots,r_{4},s_{1},\dots,s_{4}\right\}$,
which are indicated together with their degeneracies $d$. Red dots
are the compass points. The critical value $J_{{\rm rung}}^{{\rm cr}}$
for a nematic state is indicated by horizontal line. \label{fig:PD4ani}}
\end{figure}

\subsection{A ladder of $L=4$ plaquettes with anisotropic interactions}
\label{sec:aniso}

In the previous Section we studied the phase diagrams of the model
given by Eq. (\ref{eq:Ham}) with balanced XX and ZZ terms, i.e.,
they enter with exactly the same exchange interactions. It is interesting
to consider as well the present model with unbalanced interactions.
In Fig. \ref{fig:PD4ani} we show the phase diagram obtained for a
ladder of $L=4$ plaquettes in case where all the ZZ interactions are
reduced by a factor of $\gamma=0.9$ --- of course this does not affect
the symmetries that we used so far to get the block-diagonal Hamiltonian.
From the point of view of symmetries we loose the $r\leftrightarrow s$
interchange so typically the degeneracies that we find will be reduced
by a factor of 2. In the phase diagram we see that first of all the
range of stability of FMdu phase is strongly enlarged and that the
window between the two pieces of the FMu phase is enlarged too. Now
it goes above the classical threshold of $J_{{\rm rung}}=2$ and goes
lower at the high symmetry point of $J_{{\rm leg}}=J_{{\rm diag}}=1$.

The enhanced stability of the FMdu phase is clearly due to the fact
that the anisotropic interactions brake the symmetry between the $\{r_i\}$
and $\{s_{i}\}$ quantum numbers and the FMdu configuration is compatible
with such a symmetry breaking. We also notice that the quantum disorder
phases are less stable now and two of them are completely absent compared
with Fig. \ref{fig:PD4}. On the other hand, we gain two more ordered
phases, namely AF-FMd and AF which appear as bubbles instead of bubble
FMd phase which is now gone (and the wedge FMd phase is now smaller).
This shows clearly that the disordered phases were triggered by frustration
caused by the incompatibility of the XX and ZZ interactions. The nematic
phase is again unaffected.

\begin{figure}[t!]
\includegraphics[width=1\columnwidth]{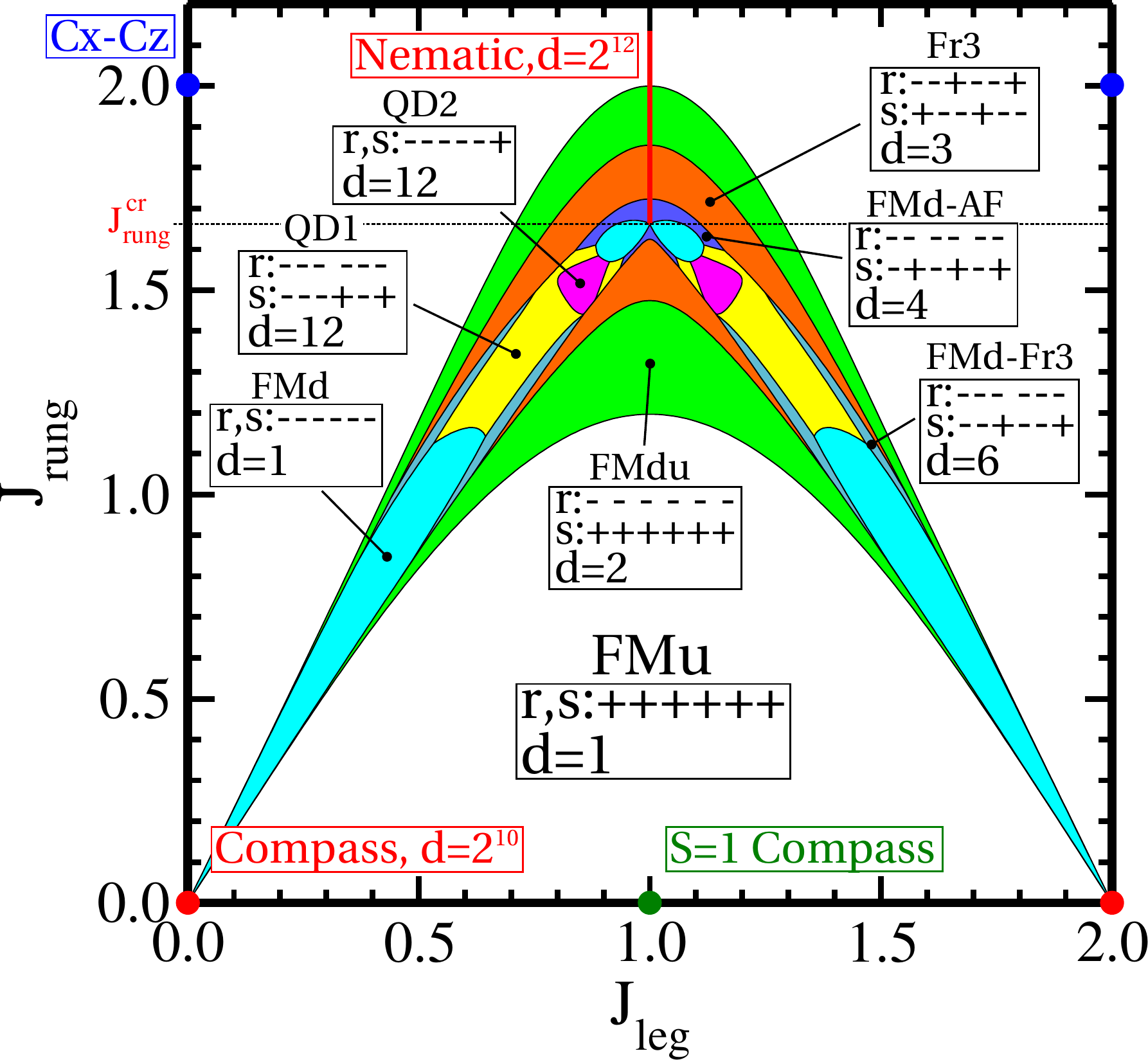}
\protect\protect\protect\caption{
Ground state phase diagram for a ladder consisting of $L=6$ plaquettes
as function of $J_{{\rm leg}}$ and $J_{{\rm rung}}$. The exchange on
diagonal bonds $J_{{\rm diag}}$ is set following  Eq. (\ref{unit}). The
ground states are shaded
in color or white and labeled according to their symmetry patterns,
$\left\{ r_{1},\dots,r_{6};s_{1},\dots,s_{6}\right\} $, which are
indicated for each phase together with their degeneracies $d$. Red
dots are the 1D compass points, blue dots indicate the simple $Cx$--$Cz$
states and green dot is $S=1$ 1D compass point. The critical value
$J_{{\rm rung}}^{{\rm cr}}$ for a nematic state is indicated by horizontal
line.}
\label{fig:PD6}
\end{figure}

\section{Generic phases for large $L$}
\label{sec:PhD2}

\subsection{Phase diagram for a ladder of $L=6$ plaquettes}
\label{sec:PD6}

We now increase the system size to identify generic phases in the phase
diagram of Fig. \ref{fig:PD4}, and to check which ones could follow
from finite-size effects. Figure \ref{fig:PD6} confirms that several
phases we have found for a short ladder of $L=4$ plaquettes reappear
for $L=6$. These are FMu, FMd, FMdu and FMd-AF phases. The disordered
phases found before are gone but there are others appearing instead.
There are also other ordered phases with a longer period $3$, namely
Fr3 (ferrimagnetic with period 3) which is stable in a large area of
the central region of the window and FMd-Fr3 which replaces former QD1
phase. Period $3$ is now allowed by the system size and we anticipate
that it is favorable from the point of view of the three-site terms
in the block-diagonal Hamiltonian so we can expect such phases whenever
$L$ is divisible by~$3$.

\begin{figure}[t!]
\includegraphics[width=1\columnwidth]{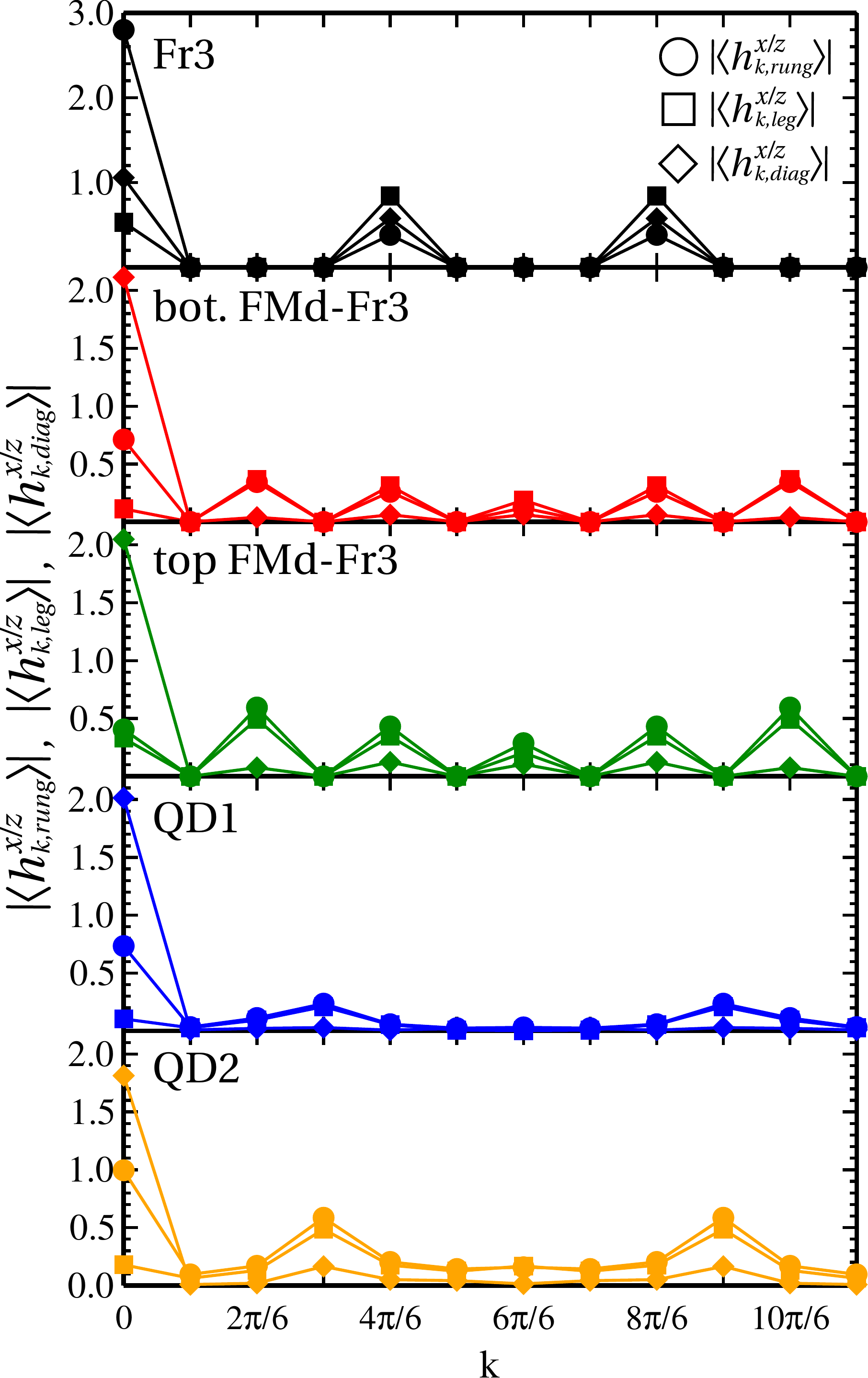}
\protect\protect\caption{
Absolute values of the Fourier transforms of the bond averages
$\langle h_{i,{\rm rung}}^{x/z}\rangle$,
$\langle h_{i,{\rm leg}}^{x/z}\rangle$, and
$\langle h_{i,{\rm diag}}^{x/z}\rangle$,
shown in Fig. \ref{fig:cnf6} in the phases FMd-AF, bot. QD4, QD3, top
QD2 and QD1. Different symbols - circles, squares, diamonds correspond
to rung, leg and diagonal bonds respectively. \label{fig:struct_6}}
\end{figure}

\begin{figure*}[t!]
\includegraphics[width=1\textwidth]{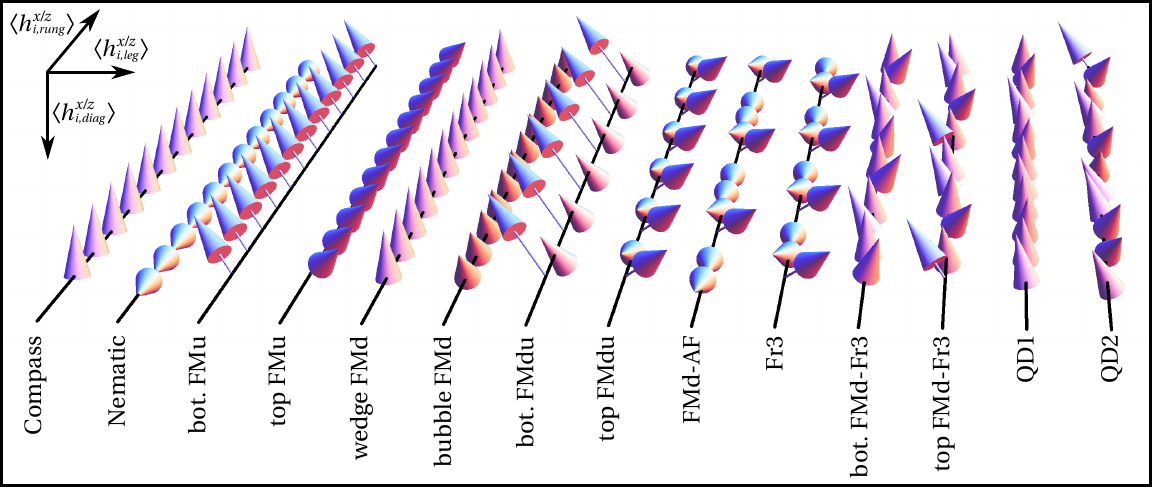}
\protect\protect\protect\caption{
Visualization of the configurations present in the phase diagram
of Fig. \ref{fig:PD6}. The solid lines indicate both system itself
and different phases or different pieces of one phase which are labeled
below. Arrows show the strength of the XX or ZZ interactions within
consecutive plaquettes of the systems such that their 3D components
correspond to ground-state average of the rung, leg and diagonal
interactions denoted as $\langle h_{i,{\rm rung}}^{x/z}\rangle$,
$\langle h_{i,{\rm leg}}^{x/z}\rangle$, and
$\langle h_{i,{\rm diag}}^{x/z}\rangle$, respectively. The reference
frame is given in the left top corner. Arrows along the solid lines
mean that only rung interactions are non-zero, vertical arrows mean
only diagonal interactions and horizontal ones only along the legs.}
\label{fig:cnf6}
\end{figure*}

Again, we find quantum disordered (QD) phases, this time QD1 and QD2.
QD1 phase takes the most of the area where FMd-AF one is stable for
$L=4$, whereas the QD2 phase appears in the form of bubbles that stem
from the bubbles of FMd phase which are now much larger than for $L=4$.
The degeneracy is twice larger in QD2 phase (and also some others)
than predicted only by a translation argument, i.e., for a label
$\vec{r}=\vec{s}=\left(-----+\right)$ we can generate only five other
distinct ground states by a translation by moving simply the $+$ spin,
suggesting $d=6$, whereas the degeneracy is \textit{de facto} $d=12$.
Now we can check the action of the $r\leftrightarrow s$ inversion; we
take the configuration $\vec{r}=\vec{s}=\left(-----+\right)$ and we get
$\vec{r}=\left(+-----\right)$ and $\vec{s}=\left(-----+\right)$ ones.
Note that quite counter-intuitively we go from the configuration with
$\vec{r}=\vec{s}$ to the one with $\vec{r}\not=\vec{s}$ keeping the
energy spectrum unchanged. Then by performing translations once again
we can get the other five configurations which explains the total
degeneracy of $d=12=2\times(1+5)$. Quite remarkably, all these changes
in the phase diagram do not affect the nematic phase which occurs for
the same parameters as before.

Figure \ref{fig:struct_6} presents the Fourier transforms of the bond
averages $\langle h_{i,{\rm rung}}^{x/z}\rangle$,
$\langle h_{i,{\rm leg}}^{x/z}\rangle$, and
$\langle h_{i,{\rm diag}}^{x/z}\rangle$ (defined by Eq. \ref{eq:struct})
in the phases with the longest periods, namely Fr3, bot. FMd-Fr3, QD3,
top FMd-Fr3, QD1 and QD2. As for $L=4$ we see that in all phases there
is a dominant ferromagnetic component at least for one type of bond
meaning large contribution at $k=0$. In case of Fr3 phase we have three
non-vanishing Fourier components for any type of bond which indicates a
three-site unit cell order. In case of mixed, FMd-Fr3 phases, top and
bottom, the number of non-vanishing components is six which corresponds
to a six-site unit cell order. As expected in the QD phases all the
Fourier components are non-vanishing indicating disorder, however for
QD1 most of them are small apart from dominant $k=0$ contribution.

Finally, in Fig. \ref{fig:cnf6} we visualize the bond averages for
different phases in the phase diagram of Fig. \ref{fig:PD6}. The phases
that were present before (for $L=4$) look here in a similar way as in
Fig. \ref{fig:cnf4} so we focus on period-$3$ and disordered phases. We
see that indeed phase Fr3 has a periodicity $3$ and every three sites
there is a strong increase in diagonal correlations which are weak
otherwise and the dominant contribution to the energy comes from the
rungs. The mixed FMd-Fr3 phase doubles the period $3$ and favors the
diagonals (or legs if $J_{{\rm leg}}>1$). Thus we see that there is a
spontaneous trimerization occurring in the phase Fr3 whereas the FMd-Fr3
one is hexamerized. The disordered QD1 phase is very weakly disordered
and similar to the wedge FMd phase or the compass one. The disorder in
the QD2 phase seems to be stronger than in the QD1 one and is
qualitatively similar to the top part of the FMd-Fr3 phase.

\subsection{Larger ladder of $L=8$ plaquettes}
\label{sec:PD8}

\begin{figure}[t!]
\includegraphics[width=1\columnwidth]{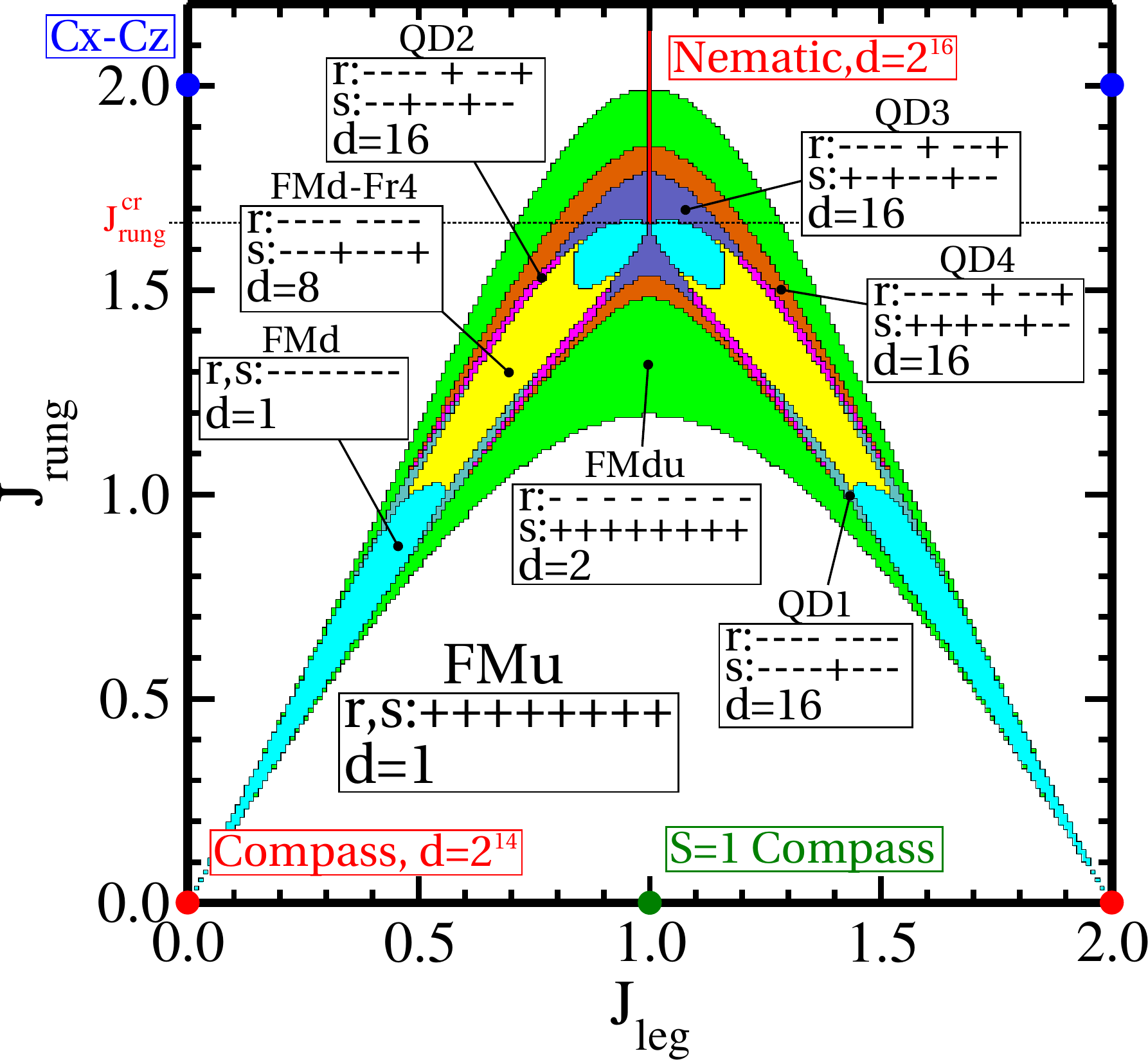}
\protect\protect\caption{
Ground state phase diagram for a ladder consisting of $L=8$ plaquettes
as function of $J_{{\rm leg}}$ and $J_{{\rm rung}}$. The exchange on
diagonal bonds $J_{{\rm diag}}$ is set following  Eq. (\ref{unit}). The
ground states are shaded
in color or white and labeled according to their symmetry patterns,
$\left\{ r_{1},\dots,r_{8};s_{1},\dots,s_{8}\right\} $, which are
indicated for each phase together with their degeneracies $d$. Red
dots are the 1D compass points, blue dots indicate the simple $Cx$--$Cz$
states and green dot is $S=1$ 1D compass point. The critical value
$J_{{\rm rung}}^{{\rm cr}}$ for a nematic state is indicated by
horizontal line.}
\label{fig:PD8}
\end{figure}

To check the generality of the phase diagrams shown in previous Sections
we explored the cases of larger plaquette ladders, $L=8$ and $L=12$. In
Fig. \ref{fig:PD8} we show the phase diagram for $L=8$. The diagram was
obtained in the following way: First the $J_{{\rm leg}}-J_{{\rm rung}}$
parameter plane was discretized as a lattice of $40\times20$ points
and in each point all $4116$ symmetry-distinct
$\left\{ r_{1},\dots,r_{L};s_{1},\dots,s_{L}\right\}$ subspaces were
searched for the ground state and then the subspaces with lowest energy
were selected as stable configurations for every lattice point.
In this way we identified eight different phases which are realized by
the $L=8$-plaquette ladder, shown in Fig. \ref{fig:PD8}. Note that
unlike in the previous cases we did not refine the boundaries between
the phases by a bisection algorithm but we used a higher-resolution
lattice of $100\times200$ points looking only at these eight subspaces
found before. Thus every pixel in the plot of Fig. \ref{fig:PD8}
symbolizes a lattice point with a color determined by the optimal
$\left\{r_{1},\dots,r_{L};s_{1},\dots,s_{L}\right\}$ configuration.

Looking at the phase diagram of Fig. \ref{fig:PD8} we observe an
overall similarity to the phase diagrams obtained for $L=4$ and $L=6$
(shown in Figs. \ref{fig:PD4} and \ref{fig:PD6}), i.e., the number of
eight distinct phases is the same as in these two cases. The main
difference is the absence of the FMd-AF phase which is here replaced by
the FMd-Fr4 one. This can be seen as a generalization of FMd-AF where
the period-$2$ AF order is replaced by Fr4 with a period-$4$. As $L=8$
is not a multiplicity of $3$, the phases with period $3$ are absent
here and the diagram resembles more the one for $L=4$. Instead of Fr3
phase one finds now two QD phases: QD2 and QD4. We remark that QD2
phase looks quite similar to Fr3 if we analyze the values of $r_i$ and
$s_i$. In addition, FMd-Fr3 phase found before at $L=6$ becomes here
QD1 phase for the same reason as above, and again certain similarity
between these phases is observed.

In fact, one may also start from a larger $L=8$ system, and one finds
that phase FMd-Fr4 is replaced at $L=6$ by two QD phases: QD1 and QD2.
These phases arise as a finite size effect but are similar to the
original FMd-Fr4 which is incompatible with the length of $L=6$.
More generally, we emphasize that the phase FMd-Fr($L/2$) is generic
and appears for any length $L=4n$, as demonstrated below by $L=12$
accessible in our analysis. Quite surprisingly, FMd-AF phase found
at $L=6$, does not survive for $L=8$ and is there replaced by QD3,
cf. Figs. \ref{fig:PD6} and \ref{fig:PD8}. This suggests that
FMd-AF phase is gradually destabilized with increasing size $L$, and
indeed it is found only is a very narrow parameter range for $L=12$,
see below.

\subsection{Largest ladder of $L=12$ plaquettes}
\label{sec:PD12}

\begin{figure}[t!]
\includegraphics[width=1\columnwidth]{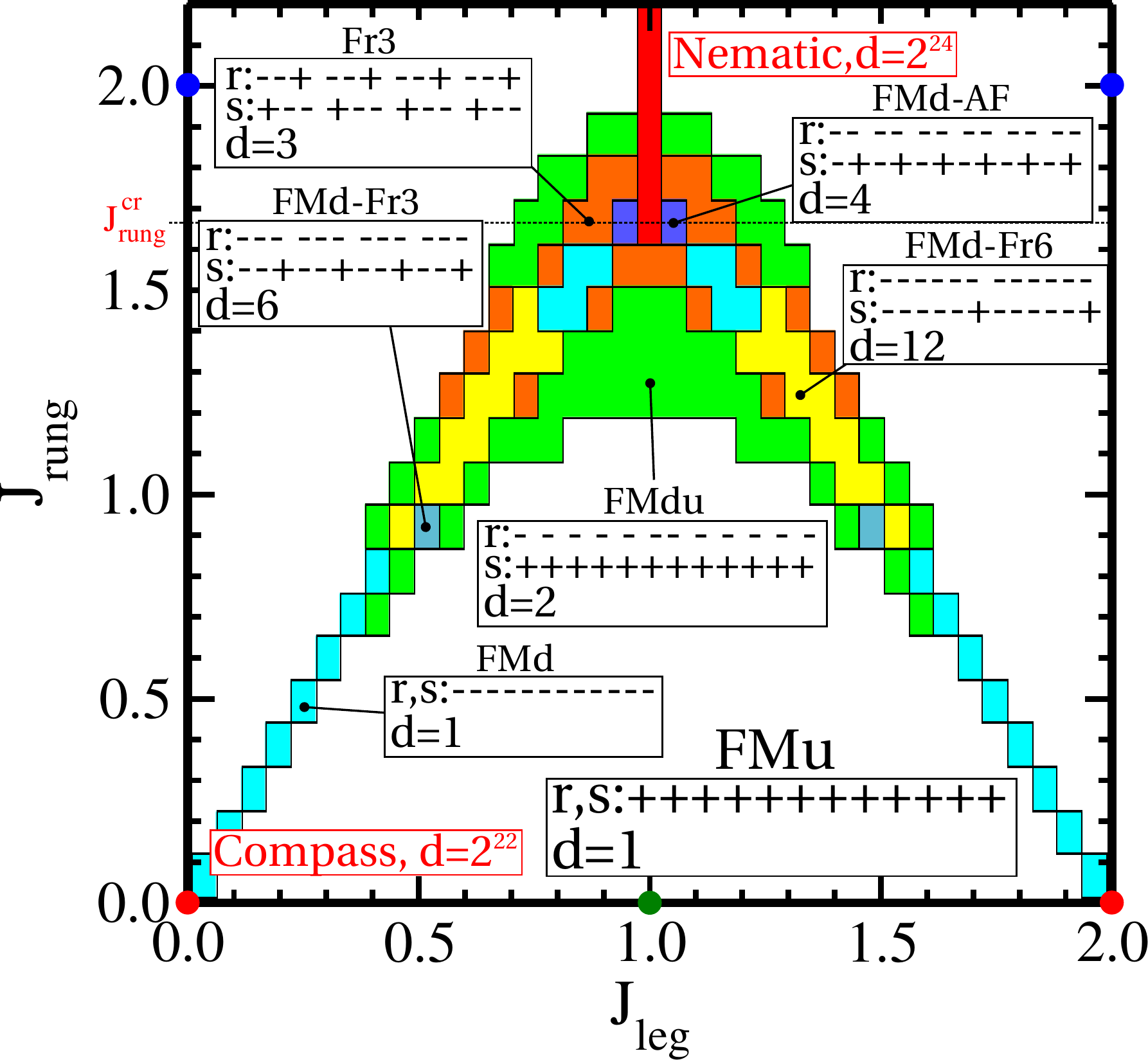}
\protect\protect\caption{
Restricted ground state phase diagram for a ladder consisting of $L=12$
plaquettes as function of $J_{{\rm leg}}$ and $J_{{\rm rung}}$
comparing stability of seven different
$\left\{r_{1},\dots,r_{12};s_{1},\dots,s_{12}\right\}$ configurations
with degeneracy $d$ indicated by color shading or white. The exchange on
diagonal bonds $J_{{\rm diag}}$ is set following  Eq. (\ref{unit}).
Red dots are the 1D compass
points, blue dots indicate the simple $Cx$--$Cz$ states and green dot
is $S=1$ 1D compass point. The critical value $J_{{\rm rung}}^{{\rm cr}}$
for a nematic state is indicated by a horizontal line.}
\label{fig:PD12}
\end{figure}

Finally we may ask what happens to the phases with period-$2$ and -$4$
in case when also period-$3$ ones are allowed by the system size $L$.
This is a case when $L$ is divisible both by $4$ and by $3$ and the
lowest possible $L$ satisfying these condition is $L=12$.
In Fig. \ref{fig:PD12} we can see a restricted phase diagram for this
case in low resolution. This follows from a lattice of $40\times20$
points in the $J_{{\rm leg}}-J_{{\rm rung}}$ plane for which we compare
the ground state energies of the eight configurations with translational
symmetry obtained for lower $L$, namely: FMd, FMu, FMdu, FMd-AF, Fr3,
FMd-Fr3, FMd-Fr4, and FMd-Fr6 phase. The latter one is another
generalization of FMd-AF phase with a longer period.
Indeed, the FMd-Fr6 phase is stable in a region between wedge and
bubble FMd phase, see the phase diagram of Fig. \ref{fig:PD12}, and
seems to be analogous of the FMd-Fr4 phase found for $L=8$ and FMd-AF
one found for $L=4$. Note that the FMd-Fr4 phase is absent here, and
the FMd-AF one appears only in two points in the phase diagram close to
the onset of the nematic phase. We anticipate that it occurs in place
of some QD phase which was not considered here.

Thus we can conclude that for general $L$ divisible by $4$ an
FMd-Fr$(L/2)$ phase appears in this part of phase diagram, with
$r_i\equiv-1$ and with most of $s_i=-1$ except for $s_{L/2}=s_L=1$. For
smaller system sizes a similar phase was less robust. For instance, the
phase diagram for $L=6$, see Fig. \ref{fig:PD6}, suggests in this
parameter range a QD1-like phase which has similar properties as the
FMd-Fr3 one --- all $r_i\equiv-1$ and almost all $s_i=-1$ except for
two sites placed irregularly. Concerning the period-3 phases, namely
Fr3 and FMd-Fr3, we see that they appear in similar positions as in the
$L=6$ phase diagram, so we argue that they are generic for $L$
divisible by $3$. Moreover, it seems that QD phases observed at smaller
system sizes appear because ordered characteristic of large $L$ cannot
yet develop.

\section{Perturbative expansion around nematic phase\label{sec:pert_nema}}

The nematic phase is the exact eigenstate of the Hamiltonian and thus
it is a good starting point of a perturbative expansion around the
high-symmetry line Eq. (\ref{line}) characterized in Sec.
\ref{sec:line}. For high enough value of $J_{{\rm rung}}$ we start
by taking the Hamiltonian in the form of Eq. (\ref{eq:Ham2}) and
dividing it into the unperturbed part ${\cal H}_{0}$ proportional
to $J_{{\rm rung}}$ and the rest which will be treated as perturbation,
${\cal V}={\cal H}-{\cal H}_{0}$, we get,
\begin{eqnarray}
{\cal H}_{0} & = & J_{{\rm rung}}\sum_{i=1}^{L}
\left(\tau_{i,2}^{x}+\tau_{i,3}^{z}\right),\\
{\cal V} & = & \sum_{i=1}^{L}
\left(\tau_{i,2}^{z}{\cal A}_{i}+\tau_{i,3}^{x}{\cal B}_{i}\right),
\end{eqnarray}
where,
\begin{eqnarray}
{\cal A}_{i} & \!\equiv\! & J_{{\rm leg}}\!\left(
r_{i}\tau_{i,3}^{z}\!+\!\tau_{i-1,3}^{z}\right)\!
+\! J_{{\rm diag}}\left(1\!+\! r_{i}\tau_{i-1,3}^{z}\tau_{i,3}^{z}\right)\!,\\
{\cal B}_{i} & \!\equiv\! & J_{{\rm leg}}\!\left(s_{i}\tau_{i,2}^{x}\!
+\!\tau_{i+1,2}^{x}\right)\!+\! J_{{\rm diag}}\left(s_{i}\!
+\!\tau_{i,2}^{x}\tau_{i+1,2}^{x}\right).
\end{eqnarray}
Note that both ${\cal A}_i$ and ${\cal B}_i$ commute with ${\cal H}_0$
so the only terms in ${\cal V}$ that will make the excitations in the
eigenstates of ${\cal H}_{0}$ are $\tau_{i,2}^{z}$ and $\tau_{i,3}^x$
that play the role of transverse fields. In the zeroth order the ground
state is a product state of the form $|\psi_0^{(0)}\rangle=\bigotimes_i
\left|\leftarrow\right\rangle_{i,2}\otimes\left|\downarrow\right\rangle_{i,3}$,
and with ground-state energy $E_0^{(0)}=-2LJ_{{\rm rung}}$. We remark
that for $J_{{\rm leg}}=J_{{\rm diag}}$ and any $r_i$ and $s_i$
operators ${\cal A}_i$ and ${\cal B}_i$ annihilate the ground state
$|\psi_{0}^{(0)}\rangle$, i.e.,
${\cal A}_i|\psi_0^{(0)}\rangle\equiv{\cal B}_i|\psi_0^{(0)}\rangle\equiv0$,
so at $J_{{\rm leg}}=J_{{\rm diag}}$ $|\psi_{0}^{(0)}\rangle$ becomes
an exact ground state of the full Hamiltonian ${\cal H}$. Using the
textbook perturbative expansion we easily find that the first order
correction vanishes, i.e., $E_{0}^{(1)}=0$ and the second order
correction has a form of,
\begin{equation}
E_{0}^{(2)}=-\frac{(\delta J)^2}{J_{{\rm rung}}}\sum_{i=1}^{L}
\left(2+r_{i}+s_{i}\right),
\end{equation}
with $\delta J\equiv J_{{\rm leg}}-J_{{\rm diag}}$. Note that the
correction gives lowest energy for $r_{i}\equiv s_{i}\equiv1$ which
means that for large enough $J_{{\rm rung}}$ close to high symmetry
line $\delta J=0$ the optimal configuration (phase) is FMu. This
agrees with all the phase diagrams shown in previous Sections. We also
easily notice that for fixed $J_{{\rm rung}}$ the energy gap around
nematic phases closes as $(\delta J)^2$.

Now it is interesting to see what happens in higher orders of the
expansion. The next non-vanishing order is the fourth order. Doing the
expansion one gets four types of contributions to the fourth order
energy correction $E_{0}^{(4)}$ but only one is of the order of
$(\delta J)^2$ and the rest is of higher orders in $\delta J$. Since
we are interested in the neighborhood of the nematic phase, only this
lowest order contribution is relevant. We get,
\begin{equation}
E_{0}^{(4)}\!=\!-\frac{(\delta J)^2(\sigma J)^{2}}{J_{{\rm rung}}^3}
\sum_{i=1}^{L}\left(2\!-r_{i}s_{i}\!-r_{i+1}s_{i}\right)
+{\cal O}\!\left(\delta J^{3}\right)\!,
\end{equation}
with $\sigma J\equiv J_{{\rm leg}}+J_{{\rm diag}}$. This correction
contains an AF interaction term between classical spins
$r_{i}$ and $s_{i}$ so the optimal configuration for $E_{0}^{(4)}$
is FMdu. We see now that the effective fourth-order Hamiltonian for
the classical spins around the nematic phase, namely
\begin{equation}
H_{{\rm eff}}\equiv E_{0}^{(0)}+E_{0}^{(2)}+E_{0}^{(4)},
\end{equation}
describes a competition between FMu and FMdu phases. In the upper
part of the phase diagram ($J_{\rm rung}>2.0$) FMu phase wins in the
large $J_{{\rm rung}}$ limit whereas FMdu phase is stable for lower
values of $J_{{\rm rung}}$. We found that the transition point is
$J_{{\rm rung}}^{\star}\simeq1.42$ and that these two phases are the
only ones that are stable. The value of $J_{{\rm rung}}^{\star}$
however does not agree with the upper boundary $J_{{\rm rung}}=2$
between FMdu and FMu phases around $\delta J=0$, found in the phase
diagrams of Figs. \ref{fig:PD4}, \ref{fig:PD6}, \ref{fig:PD8}, and
\ref{fig:PD12}. This shows that the phase competition around the
nematic phase in the low $J_{{\rm rung}}$ regime is very complex indeed
and requires higher orders of the expansion to resolve the question of
stability.

\section{Entanglement in the ladders with $L=4$ and $L=6$ plaquettes}
\label{sec:enta}

\subsection{Dimer-dimer entanglement
\label{sub:dim-dim}}

\begin{figure}[t!]
\includegraphics[width=1\columnwidth]{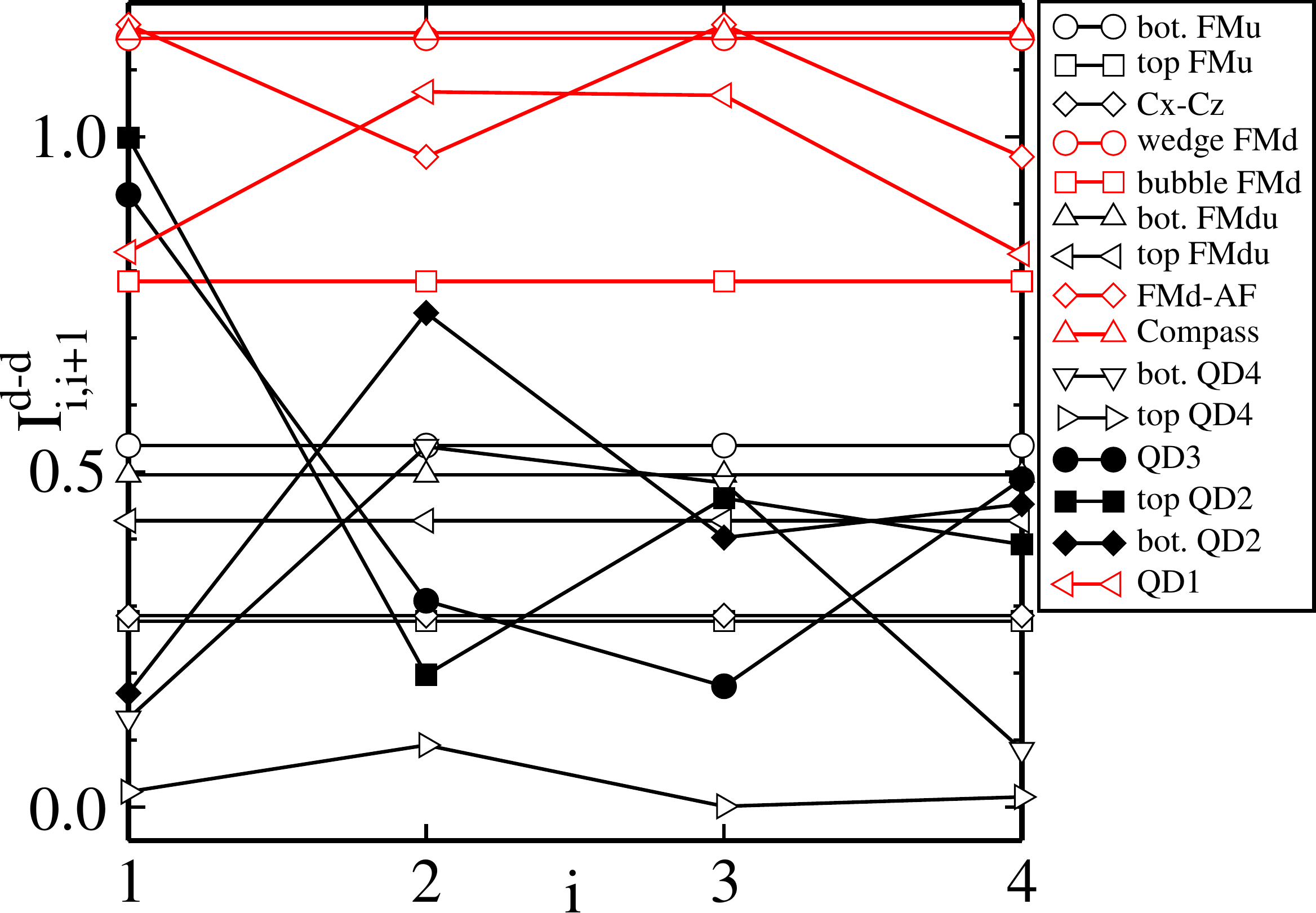}
\protect\protect\protect\caption{
Mutual information $I_{i,i+1}^{d-d}$ (\ref{info}) between the NN
dimers as a function of $i$ for a ladder of $L=4$ plaquettes obtained
in different phases and pieces of one phase shown in the phase diagram
of Fig. \ref{fig:PD4}. The (top) red lines are for phases in which
 $I_{i,i+1}^{d-d}$ is on average high ($I\simeq1$) and the black
(bottom) lines are for those in which $I_{i,i+1}^{d-d}$ is on average
low ($I\lesssim0.55$). \label{fig:mut4}}
\end{figure}

To quantify the entanglement of the states described in Figs.
\ref{fig:cnf4} and \ref{fig:cnf6} we will evaluate the mutual
information $I_{i,i+1}^{d-d}$ of the neighboring dimers (with PBCs) in
each of these states as function of the site index $i$. By a dimer we
understand pairs of transformed pseudospins $\{\tau_{i,2},\tau_{i,3}\}$
that appear in the block diagonal Hamiltonian of Eq. (\ref{eq:Ham}).
The mutual information is defined by the von Neumann entropies of the
individual dimers $i$
and $i+1$ and the pair of dimers $\left\{ i,i+1\right\}$ as follows,
\begin{equation}
I_{i,i+1}^{d-d}=S_{i}^{d}+S_{i+1}^{d}-S_{i,i+1}^{d,d},\label{info}
\end{equation}
where the von Neumann entropy $S_{A}$ of any subsystem $A$ is given
by the formula \cite{Ami08,Tho10,Lun12},
\begin{equation}
S_{A}=-\mbox{Tr}\rho_{A}\log_{2}\rho_{A},
\end{equation}
with $\rho_{A}$ being the reduced density matrix of the subsystem $A$
(i.e., the density matrix $\rho$ of the whole system is traced over all
degrees of freedom outside the subsystem $A$). Note that the subsystem
$A$ can also stand for a part of degrees of freedom in the entire
system as done in the spin-orbital systems \cite{You15}. In practice,
it is convenient to express $\rho_{A}$ in terms of ground-state
correlation functions. For instance $\rho_{i}^{d}$ for a single dimer
$i$ can be written as,
\begin{equation}
\rho_{i}^{d}=\frac{1}{2^{2}}\sum_{{\alpha,\beta=\atop 0,x,y,z}}
\tau_{i,2}^{\alpha}\tau_{i,3}^{\beta}\left\langle
\tau_{i,2}^{\alpha}\tau_{i,3}^{\beta}\right\rangle ,
\end{equation}
where $\tau_{i,p}^{0}\equiv1$ and
$\tau_{i,p}^{y}\equiv-i\tau_{i,p}^{z}\tau_{i,p}^{x}$ and where the two
$\tau$ operators in front of the average live in a local Hilbert space
of a single dimer ($\rho_{i}$ is a $4\times4$ matrix). Similarly, for
a pair of dimers we can calculate $\rho_{i,i+1}^{d,d}$ from a formula,
\begin{equation}
\rho_{i,i+1}^{d,d}\!=\!\frac{1}{2^{4}}\!\!
\sum_{{\alpha,\beta,\gamma,\delta=
\atop 0,x,y,z}}\!\!\tau_{i,2}^{\alpha}\tau_{i,3}^{\beta}\tau_{i+1,2}^{\gamma}
\tau_{i+1,3}^{\delta}\!\left\langle\!\tau_{i,2}^{\alpha}\tau_{i,3}^{\beta}
\tau_{i+1,2}^{\gamma}\tau_{i+1,3}^{\delta}\!\right\rangle\,.
\end{equation}

In Fig. \ref{fig:mut4} we show the mutual information $I_{i,i+1}^{d-d}$
(\ref{info}) for the ground states of the ladder of $L=4$ plaquettes
shown in Fig. \ref{fig:cnf4} and for ground state at the $Cx$--$Cz$
point as function of $i$. The $Cx$-$Cz$ point is taken to compare the
entanglement in the present cases to the one found for the unfrustrated
$Cx$--$Cz$ model in Ref. \cite{Brz14} which was found to be
characterized by $I_{Cx-Cz}=0.28464$. Here we recover this value working
in a different basis and we find that typically the phases not obtained
before in the unfrustrated case are more entangled. Note that the
nematic phase is not shown here because its ground state in terms of
operators $\tau_{i}$ is a classical product state and by the definition
of Eq. (\ref{info}) it has a vanishing mutual information.

\begin{figure}[t!]
\includegraphics[width=1\columnwidth]{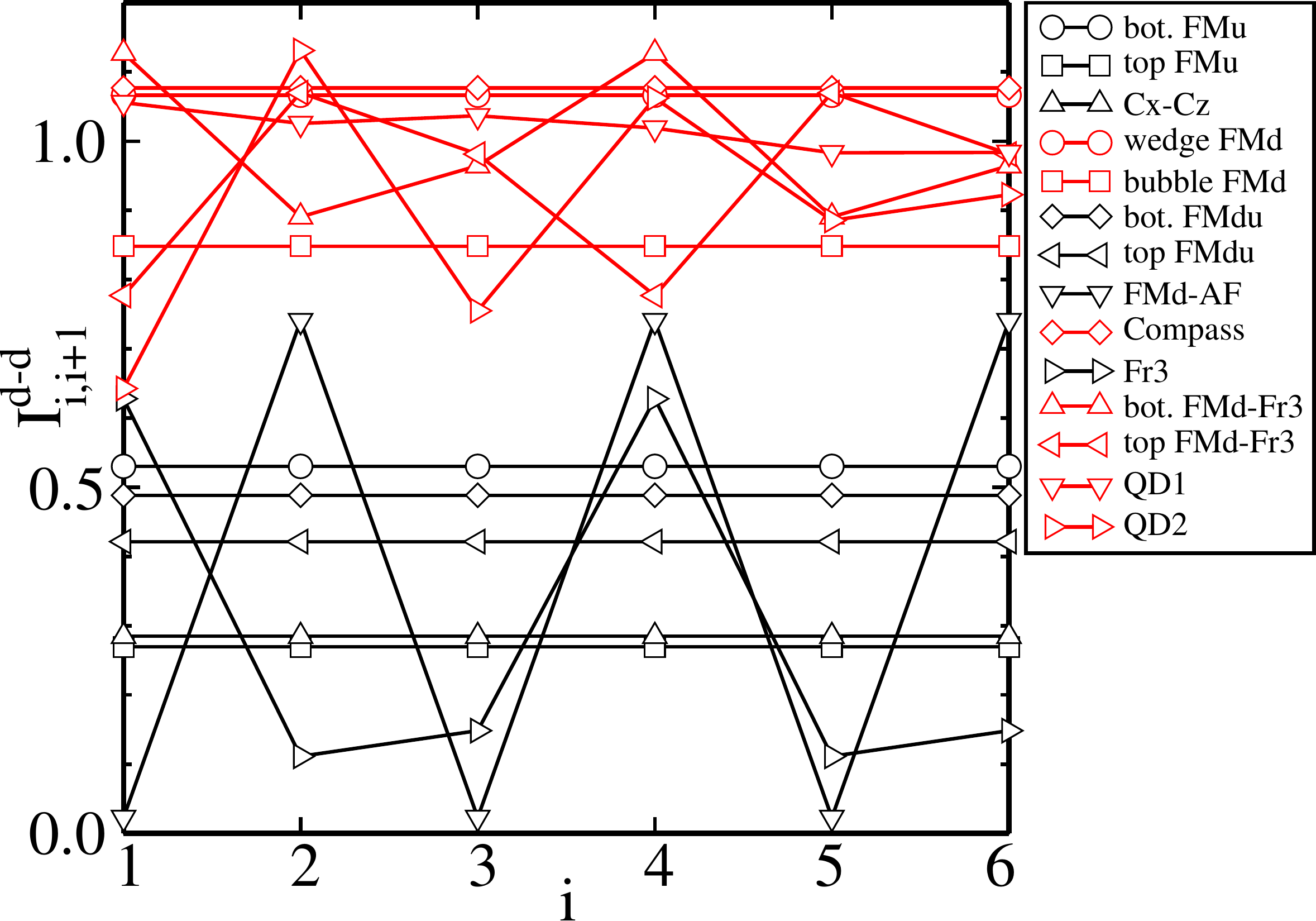}
\protect\protect\caption{Mutual information  $I_{i,i+1}^{d-d}$
(\ref{info}) between the NN dimers as a function of $i$ for a ladder of
$L=6$ plaquettes obtained in different phases and pieces of one phase
shown in the phase diagram of Fig. \ref{fig:PD6}. The red (top) lines
are for phases in which $I_{i,i+1}^{d-d}$ is on average high
($I\simeq1$) and the black (bottom) ones are for those in which
$I_{i,i+1}^{d-d}$ is on average low ($I\lesssim0.55$).}
\label{fig:mut6}
\end{figure}

In Fig. \ref{fig:mut4} we marked the phases in which $I_{i,i+1}^{d-d}$
is high ($I\simeq1.0$) on average and we find that these are both
wedge and bubble FMd phases but also FMd-AF and compass ones, together
with QD1 phase which is a tiny scrap at the interface of FMd and FMd-AF
phases (see Fig. \ref{fig:PD4}). All these phases can be seen as
an evolution of the compass phase for finite $J_{{\rm leg}}$ and
$J_{{\rm rung}}$ and indeed altogether they connect the two compass
points. The entanglement seems to decrease when $J_{{\rm rung}}$
is strongly increased, for instance in the bubble FMd phase. Quite
surprisingly the quantum disordered phase apart from QD1 do not seem
to be strongly entangled although in QD2 and QD3 phases the mutual
information $I_{i,i+1}$ (\ref{info}) can be enhanced locally to
rather high values.

As we can see from Fig. \ref{fig:mut6}, for the larger ladder of
$L=6$ plaquettes division between more and less entangled phases
seems to be clearer. Again we observe that the phases which can be
seen as continuation of the compass points in the phase diagram of Fig.
\ref{fig:PD6} are highly entangled. These are FMd, FMd-AF, QD1, QD2,
and FMd-Fr3 phases and the compass states themselves. This suggests
that the two compass points are the sources of entanglement in the
phase diagram of the extended $Cx$--$Cz$ model, and the phases that
are realized are always such that it is possible to move continuously
from one point to another keeping the entanglement high. This could
be seen as some kind of a conservation law, as if the entanglement
played a role of charge here.

Finally, we note that quite counterintuitively at the $S=1$ compass
point, i.e., $J_{{\rm rung}}=0$ and $J_{{\rm leg}}=J_{{\rm diag}}$,
the dimer-dimer mutual information is not big a for any of these two
systems and takes value of $I_{S=1}\simeq 0.45$. Naively one could
expect that it should be even higher than at the compass point
$J_{{\rm leg}}=0$ because frustration seems to be enhanced by perfectly
balanced leg and diagonal bonds. Nevertheless it is smaller and it
grows monotonously both when moving horizontally and vertically from
the $S=1$ compass point, as long as one stays within the FMu phase.

\subsection{Plaquette-plaquette entanglement\label{sub:plaq-plaq}}

\begin{figure}[t!]
\includegraphics[width=1\columnwidth]{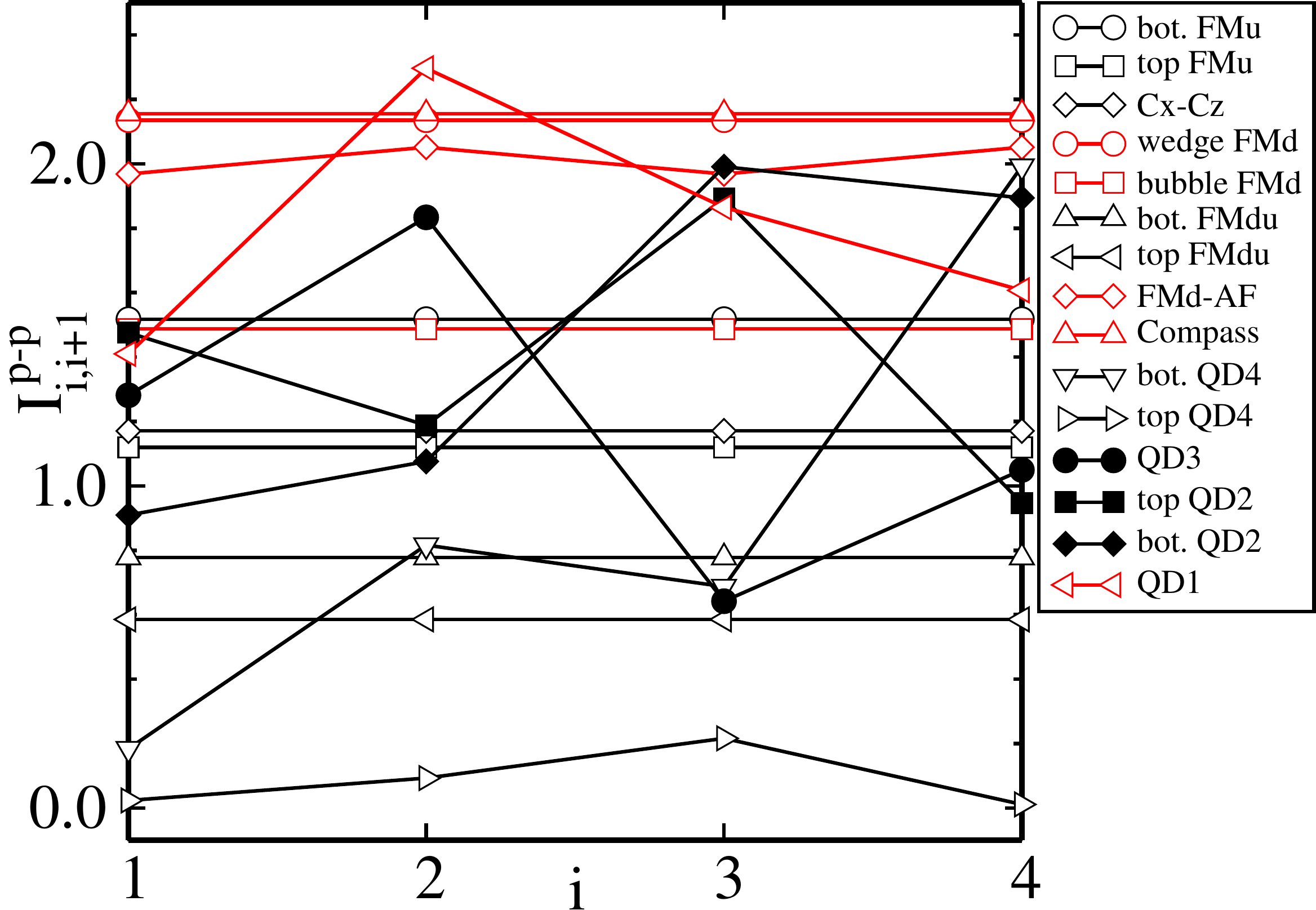}
\protect\protect\caption{Mutual information $I_{i,i+1}^{p-p}$
(\ref{info}) between the NN
plaquettes as a function of $i$ for a ladder of $L=4$ plaquettes
obtained in different phases and pieces of one phase, shown in the
phase diagram of Fig. \ref{fig:PD4}. The line colors and characters
are the same as in Fig. \ref{fig:mut4}. \label{fig:mut4ob}}
\end{figure}

Intuitively one expects that the dimer-dimer entanglement described in
the previous Subsection should be equivalent to the plaquette-plaquette
entanglement of the original ladder shown in Fig. \ref{fig:plaqlad}.
However, this is not so obvious because mutual information is not
a basis-independent quantity. The basis dependence comes from taking
the partial trace in order to obtain a reduced density matrix.
This is why we decide to examine the question of plaquette-plaquette
entanglement in the initial, physical basis. Similarly as before we                                        define the mutual information of the NN plaquettes as,
\begin{equation}
I_{i,i+1}^{p-p}=S_{i}^{p}+S_{i+1}^{p}-S_{i,i+1}^{p,p},\label{info2}
\end{equation}
where by a plaquette $i$ we understand a subsystem composed of four
initial spins $\{X_{i,p},Y_{i,p},Z_{i,p}\}$ with $p=1,2,3,4$. To
calculate this quantity we express the relevant reduce density matrices
in terms of correlation function as,
\begin{equation}
\rho_{i}^{p}=\frac{1}{2^{4}}\sum_{{A,B,C,D=
\atop 1,X,Y,Z}}A_{i,1}B_{i,2}C_{i,3}D_{i,4}
\left\langle A_{i,1}B_{i,2}C_{i,3}D_{i,4}\right\rangle ,
\end{equation}
for a single plaquette and for two plaquettes the expression is
analogous but with eight Pauli operators. Our ground state is expressed
in terms of spins $\tau_{i,2}$ and $\tau_{i,3}$ so we need to use the
transformations of Eqs. (\ref{eq:Xtrans}) and (\ref{eq:Ztrans}). The
sum contains $4^4$ terms but because of the fixed parities $r_{i}$ and
$s_i$ only 32 ground-state averages are non-zero. One has to be careful
with the signs because operators under average contain $r_{i}^{\star}$
and $s_{i}^{\star}$ which anticommute not only with $r_{i}$ and
$s_{i}$ but also with $\tau_{i,3}^{z}$ and $\tau_{i,3}^{x}$. Note
that the average will be non-zero only if all $r_{i}^{\star}$ and
$s_{i}^{\star}$ appear even number of times to cancel each other since
$(r_{i}^{\star})^{2}\equiv(s_{i}^{\star})^{2}\equiv1$. In case of
the reduce density matrix of a pairs of plaquettes the situation is
even more complicated. For non-neighboring plaquettes the number of
non-vanishing averages is simply $32^{2}$ because these plaquettes
can be treated 'independently'. In case of NN plaquettes this number
grows to $2\times32^{2}$.

\begin{figure}[t!]
\includegraphics[width=1\columnwidth]{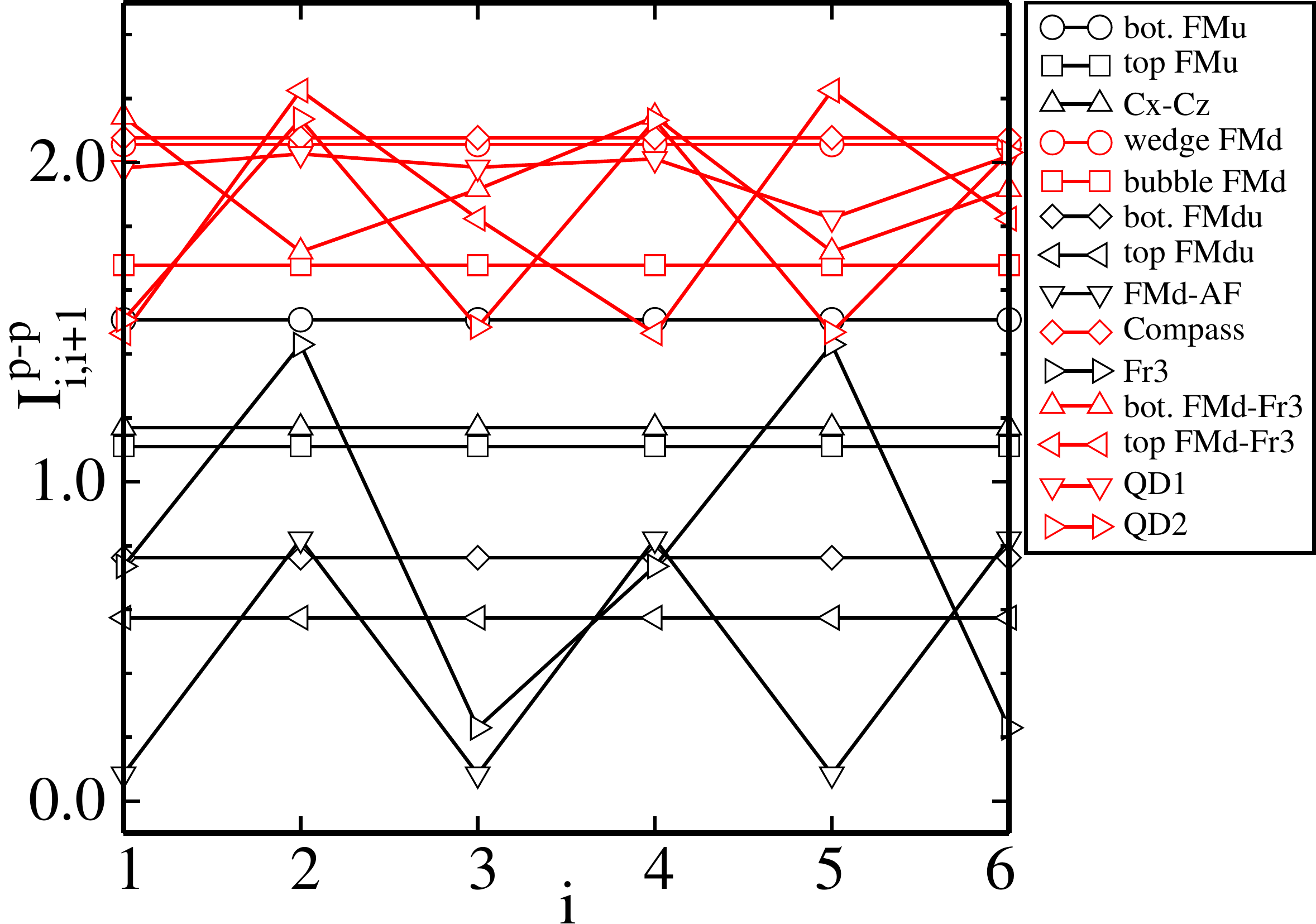}
\protect\protect\caption{
Mutual information $I_{i,i+1}^{p-p}$ (\ref{info}) between the NN
plaquettes as a function of $i$ for a ladder of $L=6$ plaquettes
obtained in different phases and pieces of one phase, shown in the
phase diagram of Fig. \ref{fig:PD6}. The line colors and characters
are the same as in Fig. \ref{fig:mut4}.}
\label{fig:mut6ob}
\end{figure}

In Figs. \ref{fig:mut4ob} and \ref{fig:mut6ob} we show the results
obtained for the mutual information of two NN plaquettes for $L=4$ and
$L=6$ for the same points in the phase diagrams as in case of dimers.
Note that the vertical scale is twice as large as for pairs of dimers
because von Neumann entropy and hence mutual information are extensive
quantities. From these plots we can draw analogical conclusions as
before because qualitatively they are similar to Figs. \ref{fig:mut4}
and \ref{fig:mut6}. This shows that the entanglement between dimers can
be treated as qualitatively equivalent to the entanglement of initial
plaquettes. Consequently, the nematic phase exhibits no
plaquette-plaquette entanglement.

\section{Summary and conclusions}
\label{sec:summa}

We have investigated the consequences of frustration in the 1D plaquette
compass $Cx$--$Cz$ model with additional frustration due to finite
antiferromagnetic exchange on plaquette diagonals. We have used a
systematic approach based on the symmetry properties and demonstrated
that different possible ground states of this model are characterized
by the eigenvalues of the local parity operators being the symmetries
of the model. A similar approach was used before for the 2D compass
model \cite{Brz10} and the 1D plaquette $Cx$--$Cz$ compass model
\cite{Brz14}, but in these cases the ground state was always found in
the simplest possible subspace where all the symmetry eigenvalues
(parities) are positive. Here we have seen that due to frustration
caused by next nearest neighbor compass interactions along diagonal
bonds in the ladder, there is a window in the parameter space
$\{J_{\rm leg},J_{\rm rung}\}$ where such a highly symmetric ground
state, called here FMu, becomes unstable and one finds instead more
exotic phase patterns containing negative parities. Examples of such
phases are an anti-FMu configuration, namely FMd where all the parities
are negative, or various configurations where the parities alternate.
Furthermore, we have shown that some of these states can be stable for
arbitrary system size whereas others are strongly related to small
system sizes, i.e., ladders of $L=4$ or $L=6$ plaquettes.

We argue that the window of exotic phases stems from the frustration
that one already finds for a single plaquette with ZZ interactions
only. Especially, the window for perfectly balanced XX and ZZ bonds
never goes above $J_{{\rm rung}}=2$, which is a value predicted by
a single plaquette study, and its shoulders cover the phase transition
lines encountered for a single plaquette. We have shown, however,
that it can become broader (exceeding $J_{{\rm rung}}=2$) when the
anisotropy between XX and ZZ bonds is introduced. We note that in
any case the window of exotic phases in the phase diagram connects
two points in the phase diagram where the model is equivalent to the
1D compass model, namely $J_{{\rm rung}}=0$ and $J_{{\rm leg}}=0(2)$,
with $J_{{\rm diag}}$ given by Eq. (\ref{unit}). It suggests that
these phases are a continuation of the degenerate manifold of ground
states of this peculiar 1D model \cite{Eri09,Liu12}.

One could expect that the ground states always respect the symmetries
of the initial ladder:
(i) translational ones and
(ii) the $r\leftrightarrow s$ interchange invariance.
However, we have observed that for ladders of up to $L=12$
plaquettes, the ground states exhibit lower symmetry than that of the
initial ladder in the highly frustrated window in the phase diagram,
and this manifests itself by ordered states with a longer period,
i.e., spontaneous multimerization. In particular, we have obtained four
phases that seem to be stable for any even $L$, two of them (FMu and
FMd) preserve full translational invariance while one (FMdu) is
dimerized and one (FMd-AF) is tetramerized. We note that a similar
dimerization due to purely quantum fluctuations was recently found in
the Kumar-Heisenberg model \cite{gang4}, but in contrast to it here
the transition to the dimerized state is discontinuous. In addition,
we have found that for a system size $L$ being a multiplicity of three
(and even), two other configurations can be stabilized, one with a unit
cell of three --- Fr3 --- and for the cell of six sites --- FMd-Fr3.
These are examples of trimerized and hexamerized ground states. We
argue that the trimmers are compatible with the three-site interactions
in the effective Hamiltonian that stem from the diagonal bonds. Apart
from these rather regular phases we have also found configurations
that we called quantum disordered in the sense that their periodicity
was equal to the system length $2L$. These phases were different for
$L=4$ and $L=6$, and we suggest that they occur only in so small
systems while they are gradually destabilized by ordered phases with
longer periods when the system size increases.

Finally, for high enough $J_{{\rm rung}}$, here
$J_{{\rm rung}}\geq J_{{\rm rung}}^{{\rm cr}}\simeq1.664$, and maximal
frustration of leg and diagonal interactions, i.e.,
$J_{{\rm leg}}=J_{{\rm diag}}=1$, we have observed a nematic phase from
$L=4$ to $L=12$ and in the anisotropic ladder of $L=4$ plaquettes as
well. Nematicity here means that the state is an eigenstate of all rung
bonds of the ladder optimizing their energies but in the same time it
is an eigenstate with zero energy of the ladder Hamiltonian without
rungs. This means that effectively only the rung bonds contribute to
the ground-state energy.
We note that this state is similar to the nematic state of the 2D
compass model \cite{Dou05,Dor05}. This observation together with the
1D compass points found in the phase diagrams makes us claim that the
present ladder model realizes indeed the paradigm of dimensional
crossover within the class of compass models.

Quite surprisingly, the vertical onset of the nematic phase is always
well within the window of exotic phases, below the classical boarder of
$J_{{\rm rung}}=2$ where the window typically closes. In this area of
the phase diagram there is clearly a very subtle competition of several
energy scales that results in what we call bubble phases being attached
to the line of nematic phase. These bubbles seem to be generic as they
are present for $L=4$, 6, and 8, as well as in the anisotropic ladder
of $L=4$ plaquettes. In absence of anisotropy these are pieces of the
FMd phase which is typically stable in the wedges touching the compass
points.

The nematic phase is characterized by a macroscopic degeneracy of
$d=2^{2L}$ related to the fact that the values of local parity
operators do not affect the energy here and there are $2L$ of them.
But in contrast to the compass model on the checkerboard lattice
\cite{Nas12}, the high degeneracy in the nematic phase and at the
compass points is not removed by quantum fluctuations. The nematic
phase appears as a highly singular part of the plaquette model where
the continuum of states is squeezed to one point.
By a perturbative expansion around the nematic phase we have
demonstrated a competition between FMu and FMdu phases. FMu (FMdu)
phase wins in the large (small) $J_{{\rm rung}}$ limit. In all phase
diagrams we have identified a special point of 1D $S=1$ compass model
which gives a gapped and unique disordered ground state at
$J_{\rm rung}=0$ and $J_{\rm diag}=J_{\rm leg}$.

As stated before, we argue that the window of exotic phases connecting
the two compass points in the phase diagram is an evolution of the
manifold of compass ground states for finite $J_{{\rm rung}}$. Quite
remarkably, this evolution seems to preserve the entanglement that
manifests itself by a high mutual information of the neighboring dimers
in the effective block-diagonal Hamiltonian or, equivalently, high
mutual information of the neighboring plaquettes in the original ladder
Hamiltonian. The mutual information
seems to be the highest at the compass points whereas it is low in the
most prolific FMu phase (including the ground state of the $Cx$--$Cz$
model \cite{Brz14}). However, in the window of exotic phases the
entanglement is typically high and comparable to the compass points.
We argue that this high entanglement originates from high frustration
that can be, to some extent, understood in terms of a single-plaquette
study.

Summarizing, we would like to emphasize that the results presented for
larger systems of $L=8$ and $L=12$ plaquettes provide enough information
to anticipate the possible phases for any $L$. We have shown that phases
FMd, FMu, FMdu appear always, as well as FMd-Fr($L/2$) for $L$ divisible
by 4, while in other cases when $L$ is divisible by 3 phases Fr3 and
FMd-Fr3 appear instead, together with a similar QD1 phase. We have
observed that quantum disordered phases appear usually when the system
size $L$ is not compatible with the actual unit cell of an ordered
phase, favored otherwise in a given range of parameters. We suggest
that the obtained phase diagram for $L=12$ plaguettes is already quite
close to the ultimate phase diagram in the thermodynamic limit.
The presented analysis of so complex phase diagrams proves that the
model is challenging enough to address the relevant phases by carrying
out an extensive density matrix renormalization group study for
relatively large system size $L$ at the level of the original ladder
Hamiltonian, as probing of all $\{r_i,s_i\}$ configurations is too
demanding beyond the system sizes considered here.

\acknowledgments

We thank Jacek Dziarmaga for helpful comments. W.B.~received funding
from the European Union's Horizon 2020 Research and Innovation
Programme under the Marie Sklodowska-Curie grant agreement No. 655515.
We kindly acknowledge support by Narodowe Centrum Nauki (NCN, National
Science Center) under Project No.~2012/04/A/ST3/00331.

\appendix*

\section{Different form of the $Cx$--$Cz$ Hamiltonian\label{sec:appA}}

For a simple $Cx$--$Cz$ Hamiltonian of Ref. \cite{Brz14} we used the
following transformation to $\sigma_{i,2/3}^{x/z}$ Pauli operators
to get its block-diagonal form,
\begin{eqnarray}
X_{i,1} & = & r_{i}^{\star},\nonumber \\
X_{i,2} & = & r_{i}^{\star}s_{i}\sigma_{i,2}^{x}\sigma_{i,3}^{x}
\left(\sigma_{i+1,2}^{x}\right),\nonumber \\
X_{i,3} & = & r_{i}^{\star}s_{i}\sigma_{i,3}^{x}
\left(\sigma_{i+1,2}^{x}\right),\nonumber \\
X_{i,4} & = & r_{i}^{\star}s_{i}\left(\sigma_{i+1,2}^{x}\right),
\end{eqnarray}
and
\begin{eqnarray}
Z_{i,1} & = & s_{i}^{\star}\sigma_{i,3}^{z},\nonumber \\
Z_{i,2} & = & s_{i-1}^{\star}
\left(\sigma_{i-1,3}^{z}\right)\sigma_{i,2}^{z},\nonumber \\
Z_{i,3} & = & s_{i-1}^{\star}r_{i}
\left(\sigma_{i-1,3}^{z}\right)\sigma_{i,2}^{z}\sigma_{i,3}^{z},\nonumber \\
Z_{i,4} & = & s_{i}^{\star}.
\end{eqnarray}
This gives in the $Cx$--$Cz$ Hamiltonian in the linear-cubic form
presented in Ref. \cite{Brz14},
\begin{eqnarray}
&\!\!&\!{\cal H}\left(J_{\rm diag}=0\right)=\nonumber\\
\!& &\!\sum_{i=1}^{L}\left\{ \left(J_{\rm leg}\sigma_{i,2}^{z}
+J_{\rm rung}\sigma_{i,3}^{z}\right)\!+\!\left(J_{\rm rung}\sigma_{i,2}^{x}
+J_{\rm leg}\sigma_{i,3}^{x}\right)\right.\nonumber \\
\!&+&\!\left. J_{\rm leg}r_{i}\sigma_{i-1,3}^{z}\left(
\sigma_{i,2}^{z}\sigma_{i,3}^{z}\right)\!+\! J_{\rm leg}s_{i}\left(
\sigma_{i,2}^{x}\sigma_{i,3}^{x}\right)\sigma_{i+1,2}^{x}\right\},
\end{eqnarray}
whereas in terms of present $\tau_{i,2/3}^{x/z}$ operators we get
a linear-quadratic form of Eq. (\ref{eq:Ham2}) which in this limit
simplifies to
\begin{eqnarray}
\!\!{\cal H}\left(J_{{\rm diag}}=0\right)\! & = & \sum_{i=1}^{L}\left\{
J_{{\rm rung}}\left(\tau_{i,2}^{x}+\tau_{i,3}^{z}\right)\right.\nonumber \\
& + & \left.J_{{\rm leg}}\left(s_{i}\tau_{i,2}^{x}\tau_{i,3}^{x}
+r_{i}\tau_{i,2}^{z}\tau_{i,3}^{z}\right)\right.\nonumber \\
& + & \left.J_{{\rm leg}}\left(\tau_{i,3}^{x}\tau_{i+1,2}^{x}
+\tau_{i,3}^{z}\tau_{i+1,2}^{z}\right)\right\}\,.
\end{eqnarray}

Note that the $\tau_{i,2/3}^{x/z}$ Pauli operators are one-to-one
related to the $\sigma_{i,2/3}^{x/z}$ ones by the following identities,
\begin{eqnarray}
\tau_{i,3}^{x} & = & \sigma_{i,3}^{x}\sigma_{i+1,2}^{x},\nonumber \\
\tau_{i,2}^{x} & = & \sigma_{i,2}^{x},\nonumber \\
\tau_{i,3}^{z} & = & \sigma_{i,3}^{z},\nonumber \\
\tau_{i,2}^{z} & = & \sigma_{i,2}^{z}\sigma_{i-1,3}^{z},
\end{eqnarray}
and the backward relations are,
\begin{eqnarray}
\sigma_{i,3}^{x} & = & \tau_{i,3}^{x}\tau_{i+1,2}^{x},\nonumber \\
\sigma_{i,2}^{x} & = & \tau_{i,2}^{x},\nonumber \\
\sigma_{i,3}^{z} & = & \tau_{i,3}^{z},\nonumber \\
\sigma_{i,2}^{z} & = & \tau_{i-1,3}^{z}\tau_{i,2}^{z}.
\end{eqnarray}

\end{document}